 \documentclass[referee]{aa} 

\usepackage{epsfig}
\usepackage{alltt,color}
\usepackage{graphicx}
\usepackage{epstopdf}
\usepackage{float}
\usepackage{palatino}
\usepackage{amsmath}
\usepackage{pifont}
\usepackage{txfonts}
\usepackage{multirow}
\usepackage{appendix}

\makeatletter

\newcommand{\Rmnum}[1]{\expandafter\@slowromancap\romannumeral #1@}
\makeatother

\begin{document}

   \title{Electron acceleration by cascading reconnection in
          the solar corona}
   \subtitle{\Rmnum{1}\: Magnetic gradient and curvature effects}


\author{X. Zhou
          \inst{1,2}
          \and
          J. B\"{u}chner
          \inst{1}
          \and
          M. B\'{a}rta
          \inst{1,3}
          \and
          W. Gan
          \inst{2}
          \and
          S. Liu
          \inst{2}
          }

\institute{Max Planck Institute for Solar System Research,
              Justus-von-Liebig-Weg 3, 37077 G\"{o}ttingen\\
              \email{zhou@mps.mpg.de}
         \and
             Key Laboratory of Dark Matter and Space Astronomy, Purple Mountain Observatory, Chinese Academy of Sciences, Nanjing, 210008, China
         \and
             Astronomical Institute of Czech Academy of Sciences, Ond\v rejov, Czech Republic
                }
\date{}

\abstract
   {}
   {We investigate the electron acceleration in
    convective electric fields of cascading magnetic reconnection in a flaring
    solar corona and show the resulting hard X-ray (HXR) radiation spectra
    caused by Bremsstrahlung for the coronal source.
    }
   {We perform test particle calculation of electron motions in the framework of a
    guiding center approximation.
    The electromagnetic fields and their derivatives along electron
    trajectories are obtained by linearly interpolating the results of
    high-resolution adaptive mesh refinement (AMR) MHD simulations of cascading
    magnetic reconnection.
    Hard X-ray (HXR) spectra are calculated using an optically thin Bremsstrahlung
    model.
    }
   {Magnetic gradients and curvatures in cascading reconnection current sheet
    accelerate electrons:
    trapped in magnetic islands, precipitating to the chromosphere and ejected into the
    interplanetary space.
    The final location of an electron is determined by its initial position,
    pitch angle and velocity.
    These initial conditions also influence electron acceleration efficiency.
    Most of electrons have enhanced perpendicular energy.
    Magnetic curvature and gradient driven acceleration efficiency along the
    magnetic field increases with the increase of the magnetic field resolution.
    It was shown that for a sufficiently high resolution of the
    electromagnetic fields, adaptive mesh refinement (AMR) MHD simulations are needed.
    Trapped electrons can be accelerated to energies up to half of $MeV$,
    precipitating ones to more than $60$ $keV$ in the
    refined smaller-scale magnetic structures.
    Trapped electrons are considered to cause the observed bright spots along
    coronal mass ejection CME-trailing current sheets as well as the flare loop-top
    HXR emissions.
    Another observable effect is the locations of the precipitating energetic
    electrons with respect to the polarity inversion line (PIL).
    A stronger asymmetry of the energetic electron precipitations around the
    PIL is obtained with sufficiently well resolved
    smaller-scale magnetic structures
    than that in the coarsely resolved ones due to the asymmetry
    in the parallel magnetic curvature acceleration.
    }
  {}

\maketitle

\section{Introduction}
\label{Introduction}
Since the first recorded white light observations of solar flares
(\citealp{Carrington1859, Hodgson1859}), sophisticated ground-based and
space-born solar techniques have been introduced to investigate the physics
of the sun.
Recently space telescopes like SoHO, Yokoh, RHESSI, Hinode and SDO, have
revealed many detailed observations covering broad wavelength
ranges at a high temporal, spatial and spectral resolution.

Generally, it is accepted that the energy of solar flares comes
from stressed, non-potential, current-carrying coronal magnetic fields being
released by magnetic reconnection.
About $10$ to $50\%$ of the flare energy may be transferred to energetic
electrons and ions (e.g. \citealp{Lin&Hudson1976}).
In some cases energetic electrons alone carried away $50\%$ of the flare
energy (e.g. \citealp{Milleretal1997}), being accelerated  to energies
up to $10-100$ $MeV$ (e.g. \citealp{Aschwanden2002}).

The prime diagnostic of accelerated electrons in solar flares is the HXR radiation
they cause.
Two main components were identified in HXR light curves
: a sharply increasing component and a slowly varying one.
The sharp increase happens within 0.5-5 s after the initial flaring
(e.g. \citealp{Holmanetal2011, Zharkovaetal2011}).
This indicates that within sub-second electrons are locally accelerated
in excess of a few $MeV$.
The slowly varying component lasts as long as
flares continue, i.e., electron energization continues.

Using high-resolution imaging, HXR location in the sun has been solved with
a few arcseconds resolution.
Solar observations have shown HXR emissions from
the foot-points of flaring coronal structures.
Recently based on the classical CSHKP
(see \citealp{Priest&Forbes2002} for a review)
solar flare reconnection model
and solar flare observations near the limb of the
solar disk, HXRs were found also at the flare loop tops (e.g.,
\citealp{Masuda1994, Gordovskyyetal2010b} with Yohkoh observation).

Although a substantial progress was made in observations, it is still an open
question by which mechanisms the flare electrons are accelerated.
Mostly suggested mechanisms can be divided into three classes
(1) acceleration by direct current (DC) electric fields (see, e.g.,
\citealp{Zharkova&Gordovskyy2004, Zharkova&Gordovskyy2005a, Zharkova&Gordovskyy2005b}),
(2) stochastic acceleration (see, e.g., \citealp{Vlahos2009}) and
(3) shock acceleration (see, e.g., \citealp{Aschwanden2002, Benz2008}).
Observations show also that different flares produce different HXR
spectra changing with time and
their locations with respect to the polarity inversion line (PIL)
(e.g., \citealp{Zharkovaetal2011}).
All these features can hardly be explained by one single acceleration mechanism.
Therefore flare energetic particles are perhaps accelerated
by different mechanisms at different time and in different places
while the flares last.

In order to validate acceleration mechanism, it is appropriate to carry out
test particle calculations.
The electron acceleration in the vicinity of a single reconnection
X-point, e.g.,  was investigated by
\citealp{Zharkova&Gordovskyy2004, Zharkova&Gordovskyy2005a, Zharkova&Gordovskyy2005b, Wood&Neukirch2005, Priest&Titov1996}).
Analytic reconnection models were used as well as the results of
ideal or resistive MHD numerical simulations.
\citealp{Martens&Young1990}, e.g., used MHD simulation results
to study the particle motions and accelerations in current sheets.
In all these models the magnetic field parallel component of the DC electric field
$\vec{E} = -\vec{u} \times \vec{B} + \eta \vec{J}$
causes strong particle acceleration if only
$\eta$ was chosen appropriately.
However, the prescription of $\eta$ in the resistive MHD simulations is
usually ad hoc and arbitrary. Meanwhile in the collisionless corona,
the concept of collisional resistivity $\eta$ is largely
unapplicable.
Microphysical effects have to be taken into account.
\citealp{Silinetal2005, Buechner&Elkina2006}, e.g., have shown
that considering possible micro-turbulence strong
parallel electric fields must be confined in narrow channels of the ion
inertia scale size (see also J. B\"{u}chner and W. Daughton 2007, section 3.5 in
\cite{Buechner2007}).
Macroscopic MHD simulations, on the other hand, is better to be used to
investigate the electron acceleration in the convective electric fields
($\vec{E}=-\vec{u}\times\vec{B}$).
\citealp{Veksteinetal1997} and
\citealp{Guoetal2010}, e.g., analysed the particle acceleration in the
convective electric fields around and at a magnetic null point, respectively.

\citealp{Veksteinetal1997} used an analytically prescribed magnetic field
with an added uniform electric field in the perpendicular direction
to calculate the test particle guiding center motions near a reconnection
X-point in a 2D geometry.
They restricted the test particle orbits far away from the X-point
since the guiding center approximation breaks down in a
null-point.
They considered particle parallel acceleration due to the $\vec{E} \times \vec{B}$
drift effects
and neglected the effects of
magnetic gradients and curvatures by launching only particles
with very small initial parallel velocities and magnetic moments.
As other authors before (e.g., \citealp{Burkhartetal1990}), they found the
final kinetic energy of the most accelerated
particle is proportional to $E^{4/3}$, where $E = -\vec{u}\times\vec{B}$.
They also assessed the spectral index of the accelerated
particles as being about 1.7,
the corresponding HXR spectral index would be around 2.7 utilizing a simple relation
$\gamma_s=\delta +1$ (where $\delta$ is the electron spectral index and
$\gamma_s$ stands for the index of emitted HXR spectrum) which is valid
within the thin target model (\citealp{Datlowe&Lin1973}).

Contrary to \citealp{Veksteinetal1997}, \citealp{Guoetal2010}
took the output of a 3D MHD simulation of magnetic null point reconnection to
study the electron and proton acceleration at a 3D null point
in the convective electric field.
Every test particle is traced by solving the full equations of motion.
This is necessary since the guiding center approximation breaks
down at a magnetic null point.
They investigated the influence of the convective speed on particle
acceleration by rescaling it.
They found that all particles are more efficiently
accelerated with a larger convective speed.
Particle energy can be up to energies of the order
of $2$ $MeV$ (proton) and $3$ $keV$ (electron) from initial thermal energy of
about $200$ $eV$.
The reason is that non-adiabatic (demagnetized) particles can
easily be accelerated in the direction perpendicular to the magnetic field.
Particle have to undergo strong perpendicular drift to be substantially
accelerated.
But particle final parallel energy can still dominate its final total kinetic
energy.
Because of the much smaller gyroradius, electrons are demagnetized
in smaller regions than the protons,
protons are accelerated to higher energies than electrons.
These authors also studied the influence of the initial energy
on particle acceleration:
higher initial energies lead protons to be stronger accelerated, while the
final kinetic energy of electrons were not influenced
essentially.

In the studies of \citealp{Veksteinetal1997} and \citealp{Guoetal2010},
there were only one magnetic X- or null point.
\citealp{Kruckeretal2008} claimed that electron DC-acceleration at
only one reconnection X-point
(e.g., \citealp{Zharkova&Gordovskyy2005a, Wood&Neukirch2005})
cannot explain the huge number of accelerated
electrons inferred from HXR observations.
The possibility of particle acceleration by cascading reconnection was
mentioned firstly by \citealp{Shibata&Tanuma2001}.
They conjectured that magnetic islands could be formed at many scales
by tearing-mode instabilities of the
stretching current sheet.
Later this concept are confirmed by theoretical approaches
(e.g., \citealp{Loureiroetal2007, Uzdenskyetal2010}),
observations (e.g., \citealp{Hoshinoetal1994, Karlicky2004}),
AMR MHD simulations (e.g., \citealp{Bartaetal2011})
and particle in cell (PIC) simulations (e.g., \citealp{Karlickyetal2012}).
The electron acceleration by many reconnection sites was studied by
\citealp{Li&Lin2012} and \citealp{Gordovskyyetal2010a, Gordovskyyetal2010b}.
In their studies, however, they assumed arbitrary ad-hoc prescribed anomalous
resistivity models to reveal the accelerating fields.
As well as the number of X-points in their studies were obtained by
periodically repeating the simulation domain.

Only the particle acceleration in the convective electric fields
$\vec{E}=-\vec{u}\times\vec{B}$, however, is independent on any ad hoc
assumption about anomalous resistivity.
In order to understand its possible acceleration effects,
we use the results of AMR-MHD simulations of multiple island formations by
cascading reconnection (\citealp{Bartaetal2010, Bartaetal2011}).
Those simulations have shown that cascading reconnection forms differently
sized magnetic islands where electrons can be accelerated (see Sect.\ref{MHD-Model}).
We use two different magnetic structure resolutions to investigated the
resolution influence on electron accelerations.
We studied the electron acceleration by cascading reconnection not only near the
X-points but also in the magnetic islands
in the framework of a
guiding center approximation (\citealp{Northrop1963}, see
Sect.\ref{Method}).
The resulting HXR emissions by energetic electron non-thermal Bremsstrahlung are
derived using a optically thin Bremsstrahlung
method (\citealp{Brown1971, Tandberg-Hanssen&Emslie1988}) to compare with
flare HXR observations.
In Sect.\ref{Results}, electron acceleration dependence on initial
conditions, different acceleration factors in the parallel direction,
acceleration in different (parallel and perpendicular) direction,
as well as trajectories,  are investigated for trapped (Sect.\ref{0_Trapped}) and
precipitating (Sect.\ref{0_Fled}) electrons.
Finally the results are discussed and conclusion
are drawn in Sect.\ref{Discussion}.

%
\section{Electromagnetic fields of cascading reconnection}
\label{MHD-Model}
In this study we aim to investigate the particle acceleration in the
convective electric fields of differently resolved cascading reconnection
current sheets
trailing a flaring arcade behind an ejected flux rope
(cf. \citealp{Lin&Forbes2000}).

The fields of cascading magnetic reconnection are obtained by means of a
2.5D AMR MHD simulation (\citealp{Bartaetal2010, Bartaetal2011}).
In traditional MHD simulations, there are only uniform grid points.
Unfortunately the sub-grid physics become important when the current sheet width
and the non-idea plasma domain become thinner than the numerical grid size.
Hence, traditional coarse MHD simulation cannot study smaller-scale
processes of anticipated cascading reconnection.
In order to resolve smaller-scale magnetic structures in the thinner current
sheets, we use simulation results which can cover an as large as possible
scale range.
The high resolution AMR MHD technique allows the description of smaller-scale
magnetic structures.
For that sake the refined mesh is used when the current sheet width becomes
comparable with the initial coarse grid size.

The AMR algorithm works as follows:
If at the time-step $t+\Delta t$ some coarse grids are detected containing thin
current sheet,
then they will locally be split into sub-boxes
with $10 \times 10$ grid-points in the
sub-system.
After such refined meshes are initialized, the necessary more detailed plasma
and field values are obtained by interpolating their parent coarse system values
at the last time step ($t$).
Then the dynamics of both the newly created and the
pre-existing refined meshes are evolved in time ($t \rightarrow t+\Delta t$)
with an accordingly refined time-step.
After that the plasma and field values at the parent coarse mesh are replaced by
averaging the quantities obtained
from its corresponding refined meshes at time-step $t+\Delta t$.
The influence of the global dynamics on the refined meshes
are considered by interpolating boundary conditions in time and space.
This refinement is repeated until the whole simulation is over (see
\citealp{Bartaetal2010}).

\begin{figure}[htbp]
      \centering
          \mbox{
                \includegraphics[width=0.45\textwidth]{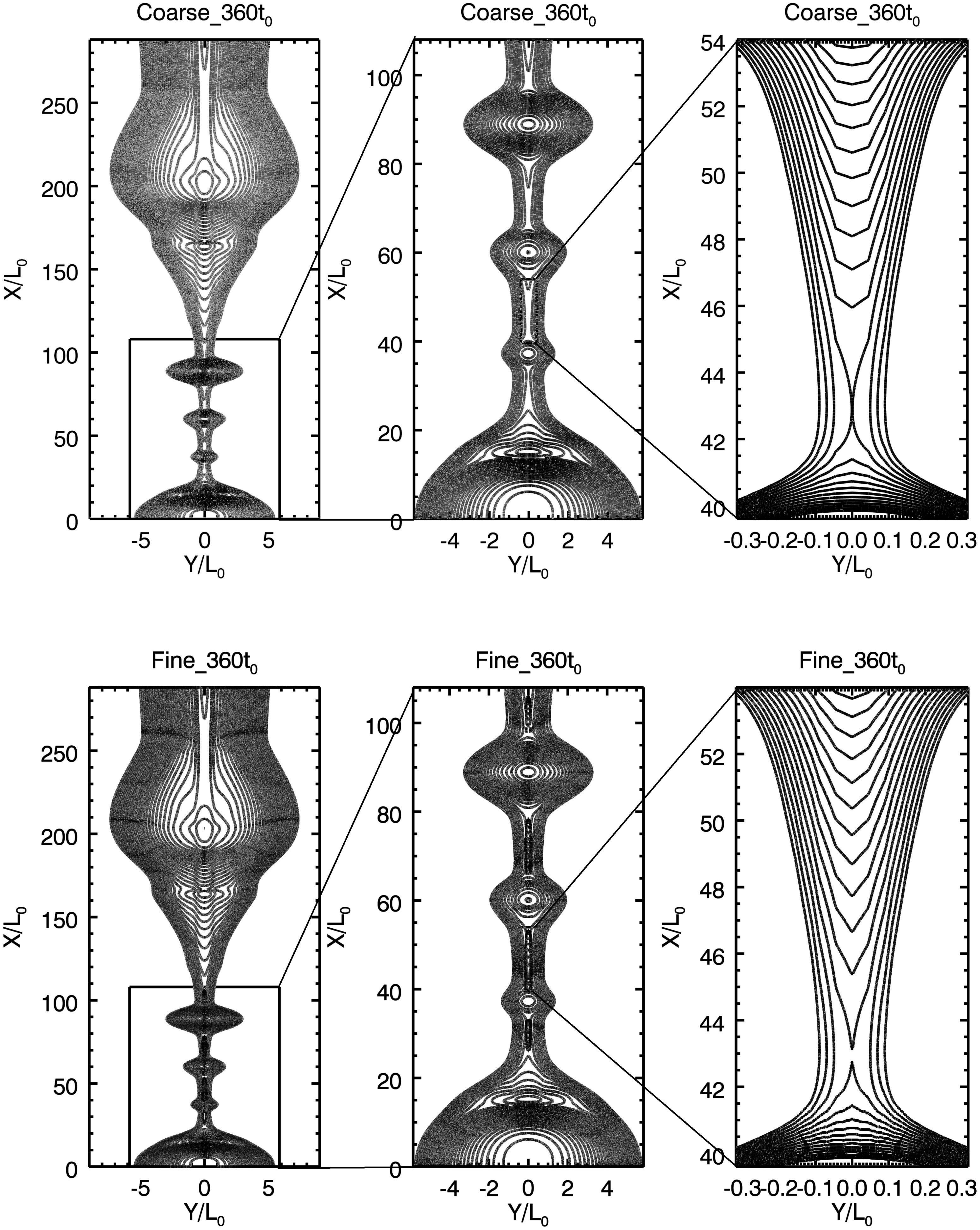}
               }
          \caption{Magnetic field lines at time $t=360$ $t_{0}$.
                  \textit{Top line}: coarsely resolved magnetic structures.
                  \textit{Bottom line}: higher resolution simulation.
                  \textit{Left} to \textit{right:} increasing zooms.
                  }
          \label{MyDomain_Zoom}
      \end{figure}

So there are two sets of electromagnetic field data
obtained by the AMR MHD simulation: one for a simulation on the coarse meshes
alone and
another with the refined meshes which provides even smaller-scale structures
of magnetic fields (see Fig.\ref{MyDomain_Zoom}).
The MHD simulation results are restricted to 2.5D, i.e.
two dimensional geometry but three dimensional plasma velocities and
magnetic fields.
This assumption is reasonable since observations have shown that the extended
solar flare arcades typically having much larger extend along the
polarity-inversion line (PIL) than across the PIL.

\begin{figure*}[htbp]
      \centering
          \mbox{
                \includegraphics[width=0.16\textwidth,viewport=193 37 435 483, clip=true]{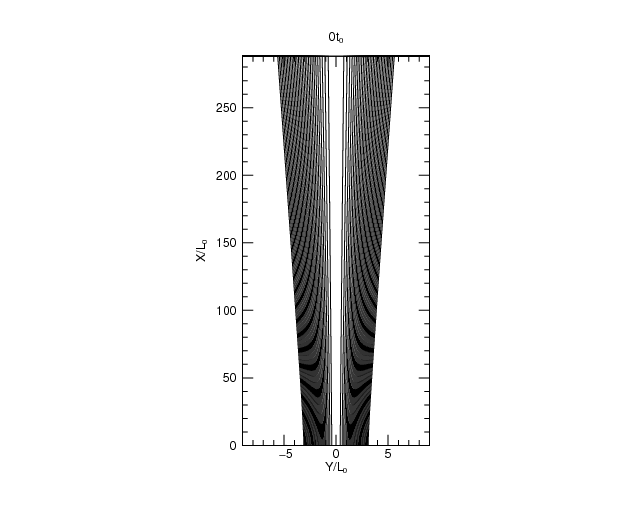}
                \includegraphics[width=0.16\textwidth,viewport=193 37 435 483, clip=true]{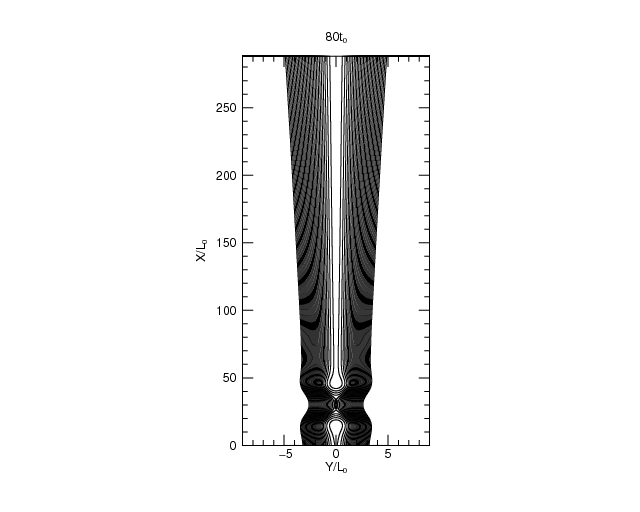}
                \includegraphics[width=0.16\textwidth,viewport=193 37 435 483, clip=true]{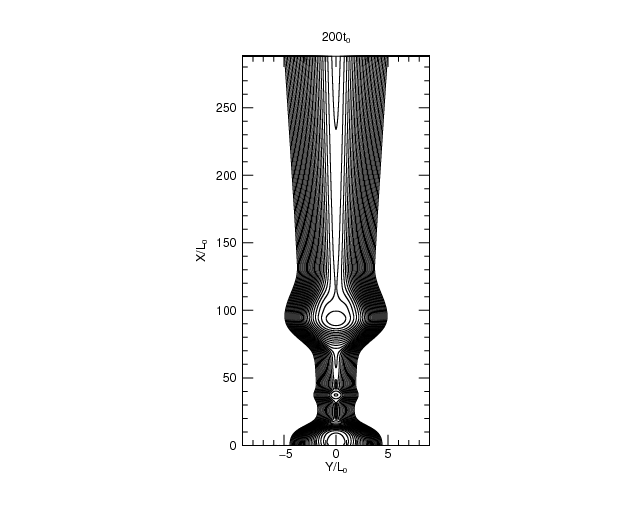}
                \includegraphics[width=0.16\textwidth,viewport=193 37 435 483, clip=true]{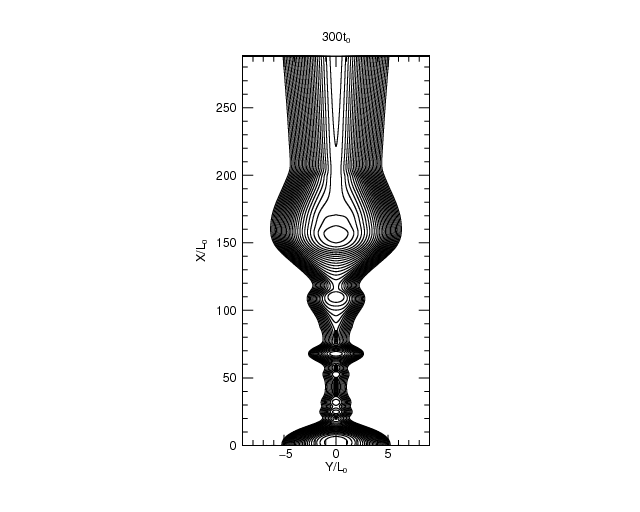}
                \includegraphics[width=0.16\textwidth,viewport=193 37 435 483, clip=true]{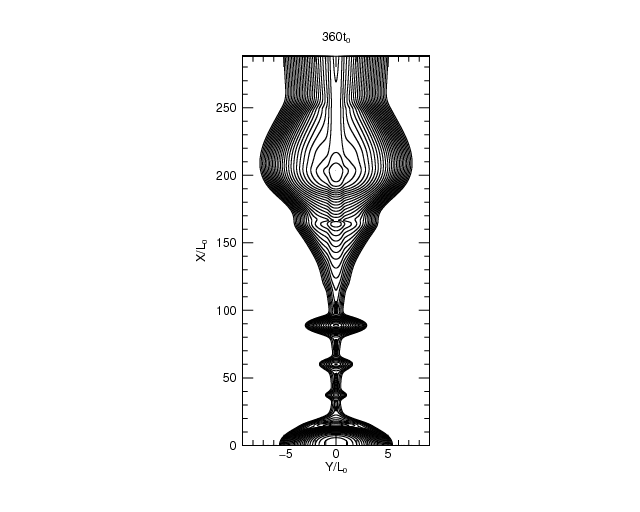}
                \includegraphics[width=0.16\textwidth,viewport=193 37 435 483, clip=true]{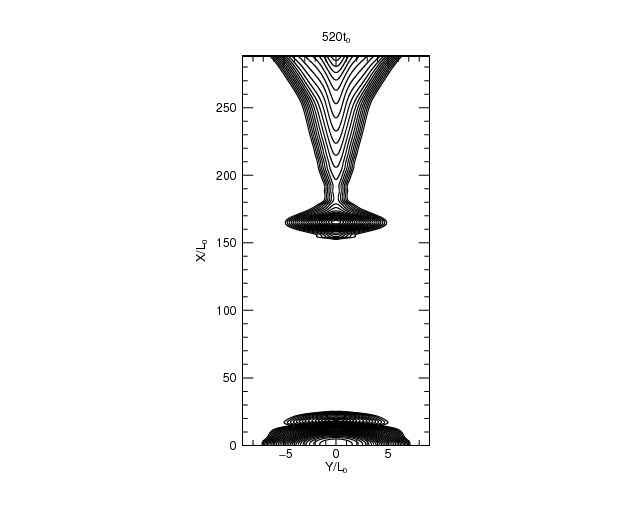}
               }
          \caption{Evolution of the in-plane magnetic field companies of cascading magnetic reconnection in the CME-trailing current sheet
                   obtained by high resolution 2.5D AMR MHD simulation.
                   Panels from \textit{left} to \textit{right} show the initial state ($t=0$ $t_{0}$),
                   primary plasmoids ($t=80$ $t_{0}$),
                   secondary plasmoids ($t=200$ $t_{0}$),
                   third stage of plasmoids ($t=300$ $t_{0}$),
                   large scale magnetic islands mature state ($t=360$ $t_{0}$)
                   and last state ($t=520$ $t_{0}$) where the erupted and disconnected magnetic field lines
                   imply the appearance of a CME.
                  }
          \label{wholeView}
      \end{figure*}

The coordinate system is shown in Fig.\ref{wholeView}: the x and y-axis are directed
along and perpendicular to the current sheet, respectively.
The current sheet center is located at $y=0$, while the z-axis is pointing
along the PIL located at ($x=0$, $y=0$).
In this direction, every value is invariant i.e. $\partial / \partial z = 0$.
The coarse resolution contains 6400 $\times$ 800 points
in the vertical (x-axis) and right half of horizontal (positive y-axis)
direction.
A mirroring boundary is used at $y=0$ for the left half
box: $\rho$, $u_{x}$, $u_{z}$, $B_{y}$, $B_{z}$ and $U$ are symmetric
while $u_{y}$, $B_{x}$ are anti-symmetric.
For the upper and right sides, free boundary conditions are used:
all quantities should satisfy the von Neumann prescription
$\partial / \partial \vec{n} = 0$ except of the normal magnetic field $B_{n}$
and the total energy density $U$. $B_{n}$ and $U$ are used
to fulfill $\nabla \cdot \vec{B} = 0$.
At the bottom, a symmetric boundary condition ($Q(-y)=Q(y)$)
is used for $\rho$, $B_{x}$, $B_{z}$, $U$ and
the anti-symmetric relation $Q(-y)=-Q(y)$ is assumed for $B_{y}$.
The plasma is always static $\vec{u} = 0$ at the bottom.

A generalized Harris-type current sheet is chosen as the initial state
of the AMR MHD simulation (\citealp{Bartaetal2010, Bartaetal2011}):
\begin{align}
\nonumber
\vec{A}(x,y,z; t=0) &= -B_{x0} \ln \Bigg[\exp \Bigg(\displaystyle\frac{y}{\omega_{cs}(x)}\Bigg)
                                       + \exp \Bigg(-\displaystyle\frac{y}{\omega_{cs}(x)}\Bigg) \Bigg] \hat{z}   \\
\nonumber
B_{z}(x,y,z; t=0) &=  B_{z0} \\
\nonumber
\rho(x,y,z; t=0) &= \rho_{0} \exp \Bigg(-\displaystyle\frac{x}{L_{G}}\Bigg)  \\
\label{InitialCS}
\end{align}
where $\omega_{cs}(x)$ (Eq.\eqref{width}) shows the characteristic width at
different height of the initial current sheet and $L_{G}$ =120 $Mm$ is the
scale hight for a fully ionized hydrogen plasma:
\begin{align}
\omega_{cs}(x)=\displaystyle\frac{d \cdot x^{2}+x+x_{0}}{x+x_{0}}
\label{width}
\end{align}
and $B_{x0}$, $B_{z0}$, $\rho_{0}$, $d$, $x_{0}$ are normalized quantities:
$B_{z0} = 0.2$, $\rho_{0} = 1.0$, $B_{0} = \sqrt{ B_{x0}^{2} + B_{z0}^{2}} =1.0$,
$d =0.003$ and $x_{0} = 20.0$.

The x and y components of the magnetic field ($B_{x}$, $B_{y}$) are obtained
from the magnetic vector potential $\vec{A}$ as $\vec{B} = \nabla \times \vec{A}$.
Note that the magnetic field strength slightly decreases via $\omega_{cs}(x)$
with height 'x' corresponding to the magnetic field in the solar corona
balancing the gravity force (Eq.\eqref{MHDEquations4}).
The initial magnetic field state is displayed in the left panel of
Fig.\ref{wholeView}.

Compressible, resistive MHD equations (Eqs.\eqref{MHDEquations1} to
\eqref{MHDEquations4}) are solved to
describe the evolution of the plasma and
magnetic fields (e.g., \citealp{Priest1984}):
\begin{align}
\displaystyle\frac{\partial \rho}{\partial t} + \nabla \cdot (\rho \vec{u})     &=      0
\label{MHDEquations1} \\
\rho \displaystyle\frac{\partial \vec{u}}{\partial t} + \rho (\vec{u} \cdot \nabla) \vec{u} &=  -\nabla p + \vec{J} \times \vec{B} + \rho \vec{g}
\label{MHDEquations2} \\
\displaystyle\frac{\partial \vec{B}}{\partial t}   &=  - \nabla \times \vec{E} = \nabla \times (\vec{u} \times \vec{B} - \eta \vec{J})
\label{MHDEquations3} \\
\displaystyle\frac{\partial U}{\partial t} + \nabla \cdot \vec{S}   &= \rho \vec{u} \cdot \vec{g}
\label{MHDEquations4}
\end{align}
where $\rho$ is the plasma density,
$\vec{u}$ plasma velocity,
$\vec{B}$ magnetic field strength,
$\vec{E}$  electric field strength,
$\eta$ resistivity,
$\vec{g}$ gravitational acceleration  at the photospheric level
and $p$ plasma pressure.
The current density $\vec{J}$, total energy density U and energy flux $\vec{S}$
are defined as:
\begin{align}
\vec{J} &= \displaystyle\frac{\nabla \times \vec{B}}{\mu_{0}} \\
U  &=  \displaystyle\frac{p}{\gamma - 1 } + \displaystyle\frac{1}{2} \rho u^{2} + \displaystyle\frac{B^{2}}{2 \mu_{0}} \\
\vec{S} &= \Bigg(U + p + \displaystyle\frac{B^{2}}{2 \mu_{0}} \Bigg) \vec{u} -
           \displaystyle\frac{\vec{u} \cdot \vec{B}}{\mu_{0}} \vec{B}
           + \displaystyle\frac{\eta}{\mu_{0}} \vec{J} \times \vec{B}
\label{EquationsS}
\end{align}
where
$\gamma_0 = \frac{5}{3}$ is the adiabatic coefficient for adiabatic condition
and
$\mu_{0}$ is the vacuum magnetic permeability.
The anomalous resistivity $\eta$ in the Ohm's law (Eq.\eqref{MHDEquations3})
and in the energy flux $\vec{S}$ (Eq.\eqref{EquationsS}) is chosen ad hoc
to describe the sub-grid-scale dissipation effects
of microphysical (kinetic) processes.
It is switched on depending on the strength of the local current-carrier drift
velocity $v_{CCD}=|J|/(e \rho)$
compared to the critical threshold velocity $v_{cr}$ (e.g., \citealp{Bartaetal2011}):
\begin{align}
\eta(\vec{r}, t)=
\left \{
             \begin{array}{cc}
             0 & |v_{CCD}| \leq  v_{cr}  \\
             C\displaystyle\frac{v_{CCD}(\vec{r}, t)-v_{cr}}{v_0} &  |v_{CCD}| >  v_{cr}
             \end{array}
\right.
\label{Resistivity}
\end{align}
\citealp{Buechner&Elkina2005, Buechner&Elkina2006} and
\citealp{Karlicky&Barta2008} have
confirmed this behaviour by means of Vlasov and PIC-code numerical simulation
and derived both critical velocity $v_{cr}$ and the coefficient $C$ in
Eq.\eqref{Resistivity}.

The AMR MHD simulation (Eqs.\eqref{MHDEquations1} to \eqref{MHDEquations4})
is carried out with normalized parameters:
the normalizing length scale (half width of the current sheet at $x=0$) is
chosen to be $L_{0} = 6.0 \times 10^{5}$ $m$,
the normalizing magnetic field is $B_{0} = 4.0 \times 10^{-2}$ $T$
and the normalizing number density is $n_{0}=1.25 \times 10^{16}$ $m^{-3}$
as well as $q_{0} = |e|$ is taken as the normalizing charge.
Other scaling parameters can be derived as:
$V_{0}=B_{0}/\sqrt{\mu_{0} n_{0} m_{0}} = 7.80 \times 10^{6} m/s$
(where $m_{0}=m_{p}$ - proton mass and $V_{0}$ is the asymptotic value
of the Alfv\'{e}n velocity at $y \rightarrow \infty$, $x=0$ and $t=0$),
The time is normalized by the Alfv\'{e}n transit time $t_{0} = L_{0}/V_{0} =
7.69 \times 10^{-2} s$.
Furthermore there are $E_{0}=V_{0} B_{0} = 3.12 \times 10^{5} V/m$ and
$\eta_{0}=\mu_{0} L_{0} V_{0} = 5.88 \times 10^{6} \Omega \cdot m$.
The asymptotic plasma beta parameter is $\beta=0.1$ at ($y \rightarrow \infty$
, $x=0$).
In the coarse resolution, the mesh sizes are $\Delta x=\Delta y=0.045$ $L_{0}$.
Hence the whole simulation domain extends over
$(0, 288) \times (-36, 36)$ $L_{0}^{2}$ in the x-y plane.

Fig.\ref{wholeView} depicts the evolution of the in-plane magnetic fields.
Total simulation is performed over 520 $t_{0}$ when a CME is ejected
through the upper boundary of the box (last panel of Fig.\ref{wholeView}).
We pick out an already fragmented current sheet at $t=360$ $t_{0}$ as the
background electromagnetic fields since at this
time step not only there are the most information of the refined smaller-scale
magnetic structures but also after that no more additional magnetic islands
are generated.
No anomalous resistivity is switched on before $t=420$ $t_{0}$.

In order to relate the electron acceleration to the
resolution of the magnetic structures,
we compare the acceleration in
the coarsely and finely resolved magnetic fields.
Fig.\ref{MyDomain_Zoom} compares the magnetic structures obtained by
the coarse (upper panels) and higher (lower panels) resolutions
at $t=360$ $t_{0}$.
From left to right, increasing zoom-levels show the details of the
magnetic structures.
The right bottom panel depict the detail of smaller-scale
magnetic structures obtained by the higher resolution.

 \section{Methods Used}
 \label{Method}
%
       \subsection{Test Particle Calculations}
        \label{Method1}

If the gyroradius ($r_{gy} = \displaystyle\frac{m v}{q B}$) and gyroperiod
($\propto 1/\omega_{gy}= \displaystyle\frac{2 \pi m}{q B}$) of the particle are
much smaller than the length scale of transverse gradients ($r_{\perp}$) and
characteristic oscillation periods ($\propto 1/\omega_{os}$) of
the ambient electromagnetic fields (i.e., $ r_{gy} / r_{\perp} \ll 1$
and $\omega_{gy} / \omega_{os} \gg 1 $),
a guiding center approximation is valid.
The motion of a magnetized charged particle can be decomposed into a
drift of its guiding center and a gyration around this center
(\citealp{Northrop1963}).

The minimum magnetic field strength obtained by the AMR MHD simulations is
$0.19$ $B_{0}$, for 10 $MeV$ energized electrons, the corresponding gyroradius
is $4.4$ $m$ only. The grid size even of the
refined mesh ($\Delta x = \Delta y = 0.0045$ $L_{0}$ $= 2.7$ km)
is much larger.
As well as, in normalized Eqs.\eqref{GC1} to \eqref{GC5} with the normalization
values shown in Sect.\ref{MHD-Model}, a coefficient
$\displaystyle\frac{m_{0} V_{0}}{q_{0} B_{0} L_{0}}$ and its reciprocal
arise in Eqs.\eqref{GC2} and \eqref{GC3}.
$\displaystyle\frac{m_{0} V_{0}}{q_{0} B_{0} L_{0}}$ corresponds to the ratio
of the particle gyroradius $\displaystyle\frac{m_{0} V_{0}}{q_{0} B_{0}}$
over the characteristic length $L_{0}$ or
the particle gyro-period $\displaystyle\frac{m_{0}}{q_{0} B_{0}}$
to the scaling time $t_{0}=L_{0}/V_{0}$.
If only $\displaystyle\frac{m_{0} V_{0}}{q_{0} B_{0} L_{0}}$ is much smaller
than unity, the guiding center approximation can be applied.
In our study, it is only of the order of $10^{-6}$.
Hence, here we use the guiding center approximation to trace each electron.

Although only 0.01\% and 0.45\% electrons can be accelerated up to
energies $> 100$ $keV$, for a high precision, a relativistic
guiding center approximation is used:
%
\begin{align}
\displaystyle\frac{d \vec{R}}{dt} = \vec{v_{D}} &+ \displaystyle\frac{(\gamma v_{\parallel})}{\gamma} \vec{b}
\label{GC1} \\
\vec{v_{D}}  = \vec{v_{E}} &+ \displaystyle\frac{m}{q} \displaystyle\frac{(\gamma v_{\parallel})^{2}}{\gamma k^{2} B} [\vec{b} \times (\vec{b} \cdot \nabla) \vec{b}]
                            + \displaystyle\frac{m}{q} \displaystyle\frac{\mu }                      {\gamma k^{2} B} [\vec{b} \times (\nabla(kB))]                         \nonumber\\&
                            + \displaystyle\frac{m}{q} \displaystyle\frac{(\gamma v_{\parallel})}    {\gamma k^{2} B} [\vec{b} \times (\vec{b} \cdot \nabla) \vec{v_{E}}]
                            + \displaystyle\frac{m}{q} \displaystyle\frac{(\gamma v_{\parallel})}    {\gamma k^{2} B} [\vec{b} \times (\vec{v_{E}} \cdot \nabla) \vec{b}]   \nonumber\\&
                            + \displaystyle\frac{m}{q} \displaystyle\frac{\gamma}                    {\gamma k^{2} B} [\vec{b} \times (\vec{v_{E}} \cdot \nabla) \vec{v_{E}}]
                            + \displaystyle\frac{1}{\gamma c^{2}} \displaystyle\frac{E_{\parallel}}  {\gamma k^{2} B} (\gamma v_{\parallel}) [\vec{b} \times \vec{v_{E}}]
\label{GC2}\\
\displaystyle\frac{d (\gamma v_{\parallel})}{dt} =&   \displaystyle\frac{q}{m} \vec{E} \cdot \vec{b}
                                - \displaystyle\frac{\mu}{\gamma}[\vec{b} \cdot \nabla(kB)]                                                     \nonumber\\&
                                + (\gamma v_{\parallel}) \vec{v_{E}} \cdot [ (\vec{b} \cdot \nabla) \vec{b} ]
                                + \gamma \vec{v_{E}} \cdot [ (\vec{v_{E}} \cdot \nabla) \vec{b}]
\label{GC3}\\
\gamma =& \sqrt{\displaystyle\frac{c^{2}+(\gamma v_{\parallel})^{2}+ 2 \mu B}{ c^{2}-v_{D}^{2}}}
\label{GC4}\\
\displaystyle\frac{d \mu}{dt} =& 0
\label{GC5}
\end{align}
here $\vec{R}$, $\vec{v_{D}}$, $v_{\parallel}$, $\gamma$ and $\vec{b}$ are the
guiding center position vector, the perpendicular drift velocity, the
velocity along the magnetic field, the relativistic factor
($\displaystyle\frac{c}{\sqrt{c^{2}-v^{2}}}$) and the magnetic field
direction unity vector
$\vec{b} = \displaystyle\frac{\vec{B}}{B}$, respectively.
In the expression for the drift velocity $\vec{v_{D}}$ in Eq.\eqref{GC2},
the term $\vec{v_{E}}$ corresponds to the local $\vec{E} \times \vec{B}$
drift velocity $\vec{v_{E}} = \displaystyle\frac{\vec{E} \times \vec{B}}{B^{2}}$.
Other terms are the magnetic curvature drift velocity and
the magnetic gradient drift velocity as well as higher order drifts.
The factor $k = \sqrt{1-\displaystyle\frac{ \vec{v_{E}}^{2} }{c^{2}}}$
relates the electromagnetic field values to the reference frame moving with the
velocity $\vec{v_{E}}$.
Finally, $\mu = \displaystyle\frac{(\gamma v_{\perp})^{2}}{2B}$ is the relativistic
magnetic moment per mass unit where $v_{\perp}$ is the particle gyration velocity
perpendicular to $\vec{B}$.
The electron energy is expressed using the relativistic $\gamma$-factor as
$E=(\gamma-1)mc^{2}$.
The set of Eqs.\eqref{GC1} to \eqref{GC5} are solved utilizing a
fourth-order Runge-Kutta scheme. The field values between the grid points are
interpolated along the electron trajectories with 2D linear interpolation.


$4.752\times10^{5}$ test electrons are initially uniformly distributed
along the current sheet ($0 < x < 108$ $L_{0}, y =0$) at 2400 points with
22 different initial velocities from
0.0 to $21.0$ $v_{th}$ and 9 different initial pitch angles from
0 to $\pi$.
Here $v_{th}$ is the electron thermal velocity for a typical coronal
temperature of $10^6$ $K$: $0.76 V_{0} \cong 6\times 10^{3}$ km/s.
Every electron is traced for up to $10 t_{0}$ ($\sim 0.769$ $s$) or until
it leaves the simulation domain, whatever happens first.
Not that this time is shorter than the time scale of essential magnetic field changes
in the MHD simulations.

       \subsection{Spectrum $\/$ Distribution function of accelerated electrons}
        \label{Method3}

To obtain the energetic electron distribution function, we use the fact that the
solar corona is practically collisionless. Hence
according to Lioville's theorem, the particle
distribution function keeps constant along the particle trajectory:
$f(E, A, \vec{r}, t) = f(E_{0}, A_{0}, \vec{r_{0}}, t_{0})$.
This allows to calculate the electron distribution function $f(E, A, \vec{r}, t)$
at the place where HXR are expected to be generated by Bremsstrahlung of
energetic electrons.

       \subsection{HXR Emission}
        \label{Method2}
Knowing the local plasma number density and electron distribution function,
the hard X-ray emissivity $I(\epsilon)$ integrated
over all contributing electrons can be calculated in the frame work of
the thin target model (\citealp{Brown1971}) as:
\begin{align}
I(\epsilon) =  \sum^{\infty}_{E>\epsilon} n(\vec{r}) v(\vec{r}) \sigma_{B}(\epsilon, E) f(E, A, \vec{r}, t)
\label{HXREquation}
\end{align}
Here $E$, $A$, $\vec{r}$ and $v(\vec{r})$ are the energy, pitch angle,
position and velocity of the electron at the time $t$,
$n(\vec{r})$ is the local plasma number density,
$\epsilon$ is the radiated photon energy,
$f(E, A, \vec{r}, t)$ is the electron distribution function at the position
of interest place and time, while $\sigma_{B}$ is the cross section of
the Bremsstrahlung process. For a simple approximation, we
take the Bethe-Heitler formula for the Bremsstrahlung cross section
(\citealp{1934RSPSA.146...83B, Brown1971}):
\begin{align}
\sigma_{B}(\epsilon, E) \propto \displaystyle\frac{1}{\epsilon E}
                                \ln \Bigg[\displaystyle\frac{1+\sqrt{1-\displaystyle\frac{\epsilon}{E}}}
                                                            {1-\sqrt{1-\displaystyle\frac{\epsilon}{E}}}\Bigg]
\label{Sigma}
\end{align}
Note that the Bethe-Heitler formula applies only to particle energies less than
$100$ $keV$.
In this investigation, for both kinds of magnetic fields resolution,
more than $99\%$ of the electrons are accelerated to energies less than $100$
$keV$, i.e. the Bethe-Heitler formula still can give a high accuracy here.

\section{Results}
\label{Results}

Depending on the locations of the simulated electrons at 10 $t_{0}$,
three groups of electrons can be identified: those trapped in the magnetic
islands; those precipitating to the chromosphere
and the ones being ejected into the interplanetary space.
There is no electron escaping from the left and right sides of the simulation
domain.
We concentrate our analysis on the trapped (Sect.\ref{0_Trapped}) and
precipitating (Sect.\ref{0_Fled}) electrons.

          \subsection{Trapped Electrons}
          \label{0_Trapped}

More than 80\% of simulated electrons are still trapped in the magnetic islands
along the current sheet by $10$ $t_{0}$. This highly dynamical and complex
magnetic field structure provides a very effective trapping mechanism
of energetic electrons for the coronal HXR sources.

\subsubsection{Acceleration dependence on initial conditions and magnetic field resolution}

\begin{figure}[htbp]
            \centering
                \mbox{
                       \includegraphics[width=0.25\textwidth]{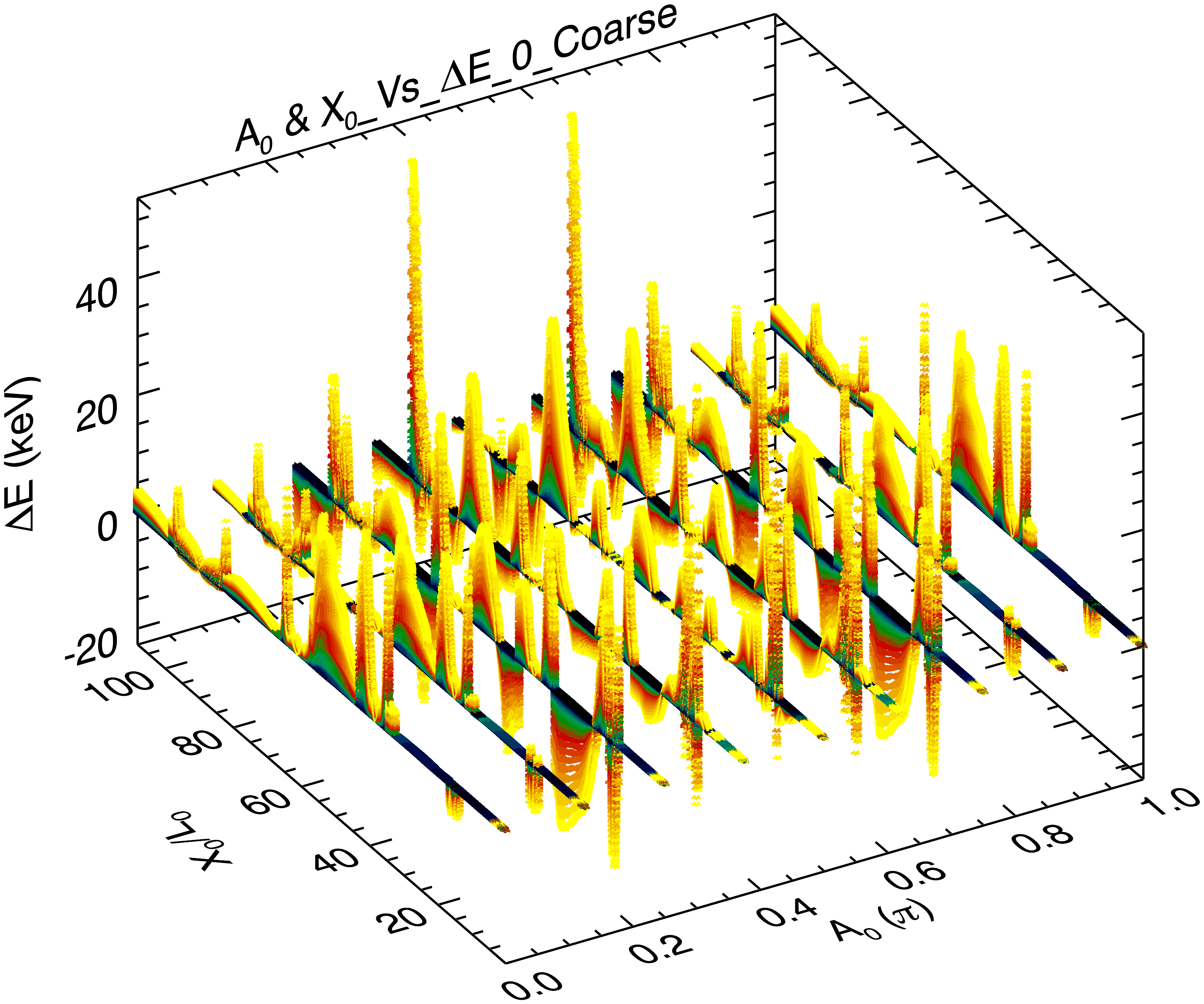}    %
                       \includegraphics[width=0.25\textwidth]{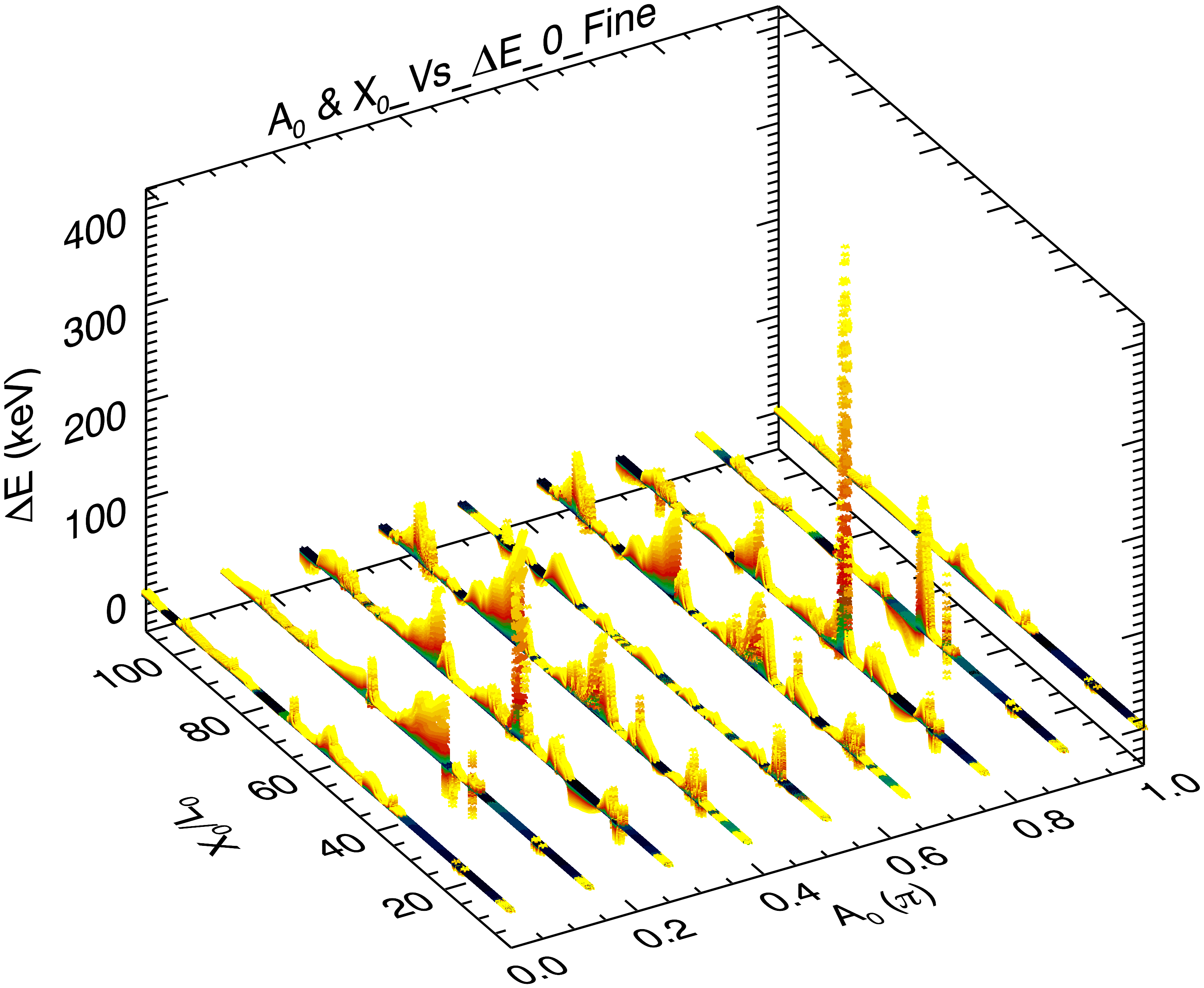}
                     }
                     \\
                \mbox{
                       \includegraphics[width=0.50\textwidth]{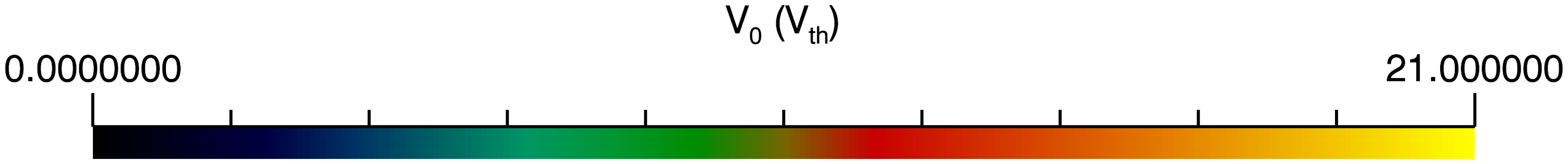}
                     }
                     \\
                \mbox{
                       \includegraphics[width=0.50\textwidth]{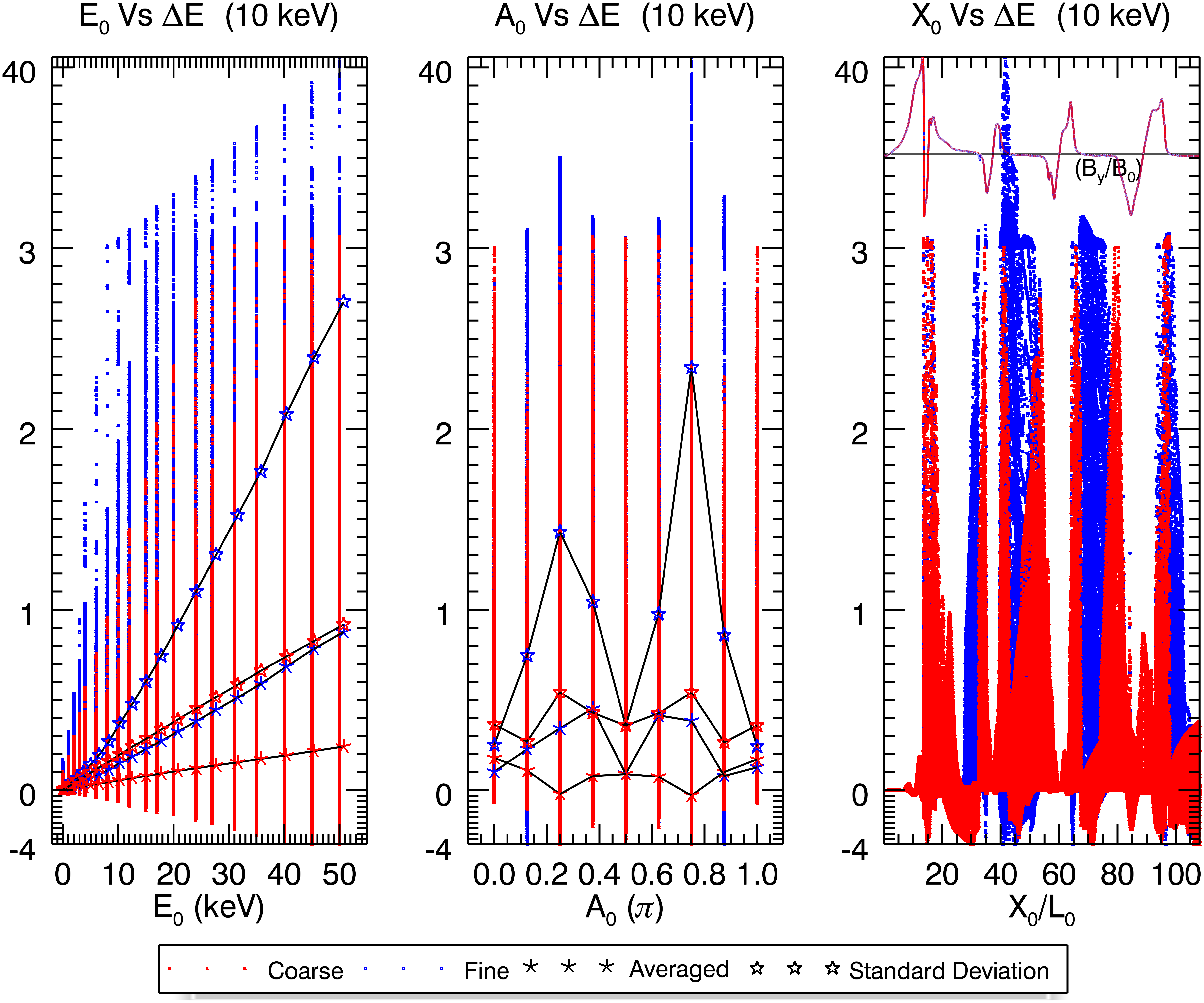}
                     }
                \caption{ Dependence of the electron kinetic energy gain ($\Delta E$) on the initial pitch angle ($A_{0}$), velocity ($v_{0}$)
                          and position ($x_{0}$) for two differently resolved magnetic field structures. Each point represents one electron.
                          The \textit{upper} panels show the 3D result with the initial velocity color coded.
                          The \textit{lower} panels show the corresponding projected results and the averaged value ($\star$ lines) and
                          standard deviation of the energy gain (\ding{73} lines) in the initial energy and pitch angle spaces.
                          $B_{y}$ along the current sheet center is depicted in the bottom-right panel.
                          The \textit{red} and \textit{blue} colors are used to distinguish the results in
                          the coarsely and finely resolved magnetic fields, respectively.
                          Note that there are three different scales in the y-axis for $\Delta E <0$, $0<\Delta E<3$ $keV$ and $\Delta E > 3$ $keV$.
                         }
                \label{0_0_Trapped}
            \end{figure}

The acceleration of electrons in the convective (or induced) electric field
$\vec{E}=-\vec{u}\times\vec{B}$ is sensitive to the initial position, velocity,
and pitch angle of injected electrons and the fine structure of the magnetic
field.
The upper panels of Fig.\ref{0_0_Trapped} depict the dependence of
the energy gain on the initial conditions and the magnetic field resolution.
The lower panels show the corresponding projected results.

In general, the electron acceleration is more efficient in magnetic fields
with better resolved small scale structures for the larger magnetic
curvatures and gradients accessible.
The maximum final kinetic energy of trapped electrons
is at most of the order of $100$ $keV$ in coarsely resolved magnetic fields,
but it can be up to $470$ $keV$ if smaller-scale magnetic structures are taken
into account, corresponding to a maximum energy gain of
$53$ $keV$ and $420$ $keV$ for the coarsely and finely resolved fields
respectively.

From left to right the bottom panels of Fig.\ref{0_0_Trapped} depict
the dependence of acceleration efficiency on the electron initial energy,
pitch angle and position, respectively.
The kinetic energy gain increases with the increase of the initial energy,
which is consistent with the results of \citealp{Guoetal2010}.
It is interesting to note that both the mean and the standard deviation of
the energy gain are roughly proportional to the initial energy and the
acceleration efficiency of the finely resolved case is about 3 times higher
than the coarse one.
The dependence of the electron energy change on the initial pitch angle and
position, however, is more or less chaotic due to the complex field structures.
Different from \citealp{Karlickyetal2004} where the betatron process dominates,
here the most energetic electron
is not associated with an initial pitch angle of $90^{\circ}$ any more.
The acceleration symmetry with respective to the $90^{\circ}$ pitch angle is
also broken when the magnetic fields are better resolved.
The lower-right panel of Fig.\ref{0_0_Trapped} also shows the magnetic field
component $B_{y}$ along the current sheet center $y=0$, which can be used to
identify the magnetic X- and O-points with $B_{y}=0$.
The most efficient acceleration appears to be associated with electrons
injected close to the X-points that contain larger magnetic gradients and
smaller magnetic curvature radii.


\subsubsection{Energy gain}
\label{Acceleration}

The guiding center approach decomposes particle energy into components
parallel and perpendicular to the magnetic field and the part associated with
the guiding center drift in the direction perpendicular to the magnetic
field.
The maximum drift velocity $v_{D}$ (Eq.\eqref{GC2}) is
1.40 $v_{th}$ and 0.97 $v_{th}$ in the coarsely and finely resolved
magnetic fields, which is negligible comparing with the other two
components.

Considering that the anomalous resistivity is not switch on
($\vec{E} \cdot \vec{b}=0$) and
$\vec{v_{D}} \cong \vec{v_{E}}$, $k \cong 1$, Eqs.\eqref{GC3}, \eqref{GC5} and
$\mu = \displaystyle\frac{(\gamma v_{\perp})^{2}}{2B}$ give:
\begin{align}
\displaystyle\frac{1}{2}\displaystyle\frac{d (\gamma v_{\parallel})^{2}}{dt} =&
                                - \mu v_{\parallel}[\vec{b} \cdot \nabla B]
                                + (\gamma v_{\parallel})^{2} \vec{v_{E}} \cdot [ (\vec{b} \cdot \nabla) \vec{b} ]
\label{Parallel_Gain}\\
\displaystyle\frac{1}{2}\displaystyle\frac{d (\gamma v_{\perp})^{2}}{dt} =& \displaystyle\frac{d \mu B}{dt} = \mu \displaystyle\frac{d B}{dt}   \nonumber\\
=& \mu v_{\parallel}[\vec{b} \cdot \nabla B]+ \mu \vec{v_{E}} \cdot \nabla B
\label{Perpendicular_Gain}
\end{align}
So the energy evolution of an electron in the guiding-center limit
is given by:
\begin{align}
\displaystyle\frac{d E_{k} }{dt}
\propto & \displaystyle\frac{d (\gamma v_{\parallel})^{2} +(\gamma v_{\perp})^{2}}{dt}
\nonumber\\
= &   \mu \vec{v_{E}} \cdot \nabla B
    + (\gamma v_{\parallel})^{2} \vec{v_{E}} \cdot [ (\vec{b} \cdot \nabla) \vec{b} ]
\label{Total_Gain}
\end{align}

\begin{figure}[htbp]
            \centering
                \mbox{
                       \includegraphics[width=0.50\textwidth]{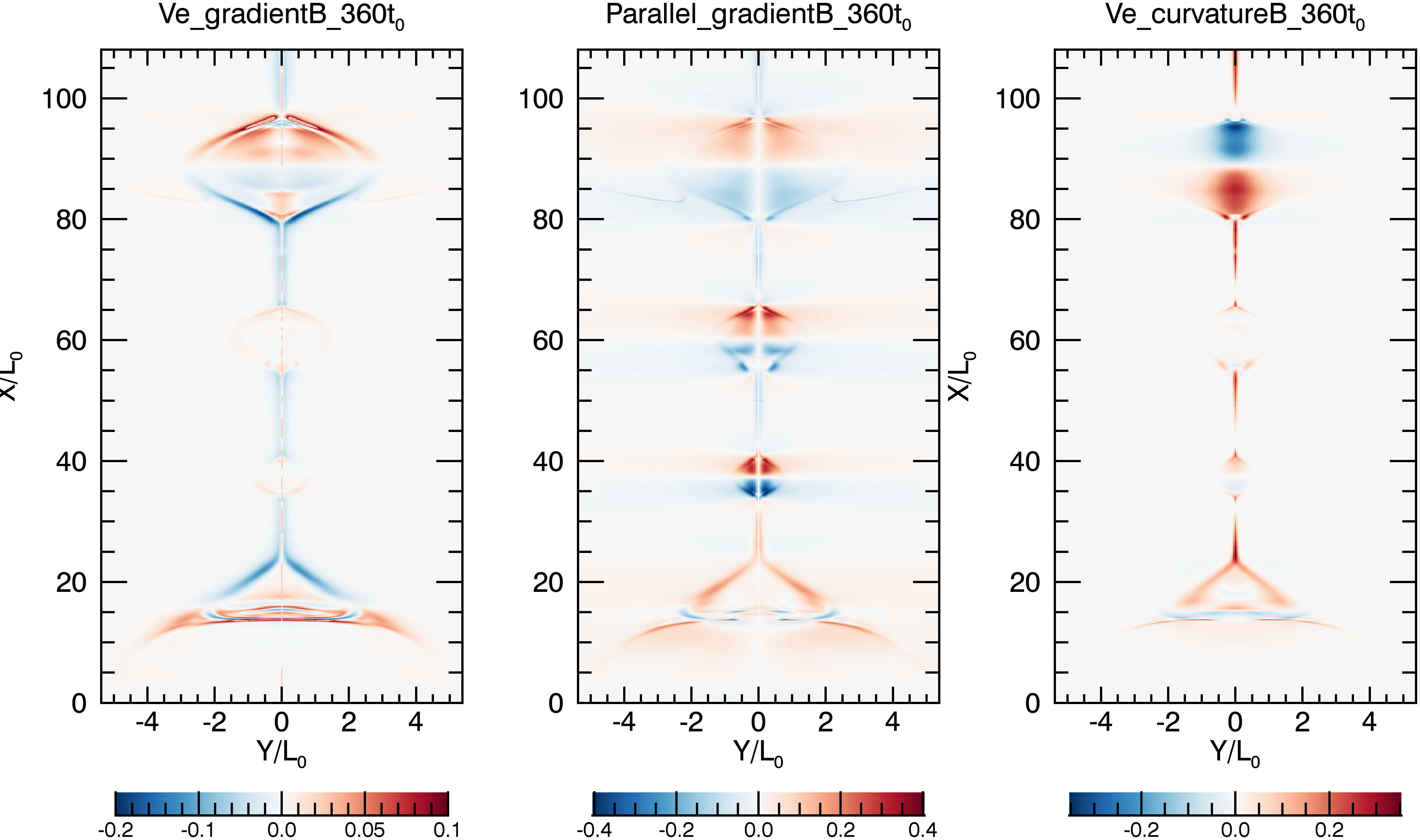}
                     }
                \caption{Spacial distributions of perpendicular (\textit{left} panel), parallel (\textit{middle} panel) gradient
                         and perpendicular curvature ((\textit{right} panel))
                         in Eqs.\eqref{Parallel_Gain} and \eqref{Perpendicular_Gain} of the coarsely resolved magnetic fields.
                         }
                \label{Acc_Factors}
            \end{figure}

Fig.\ref{Acc_Factors} exhibits the spacial distributions of the acceleration
rates in Eqs.\eqref{Parallel_Gain} and \eqref{Perpendicular_Gain}:
$(\vec{b} \cdot \nabla B)/ (2B)$ (the middle panel),
($\vec{v_{E}} \cdot \nabla B )/ (2B)$ (the left panel)
and
$\vec{v_{E}} \cdot [(\vec{b} \cdot \nabla) \vec{b}]$ (the right panel) in the
coarse calculation domain (corresponding to the top-middle panel of
Fig.\ref{MyDomain_Zoom}), note that these distributions are quite similar
between the coarsely and finely resolved magnetic fields.

Eqs.\eqref{Parallel_Gain} and \eqref{Perpendicular_Gain}
depict the parallel magnetic gradient ($\vec{b} \cdot \nabla B$) can change
both the parallel and perpendicular energies of electrons, but they cancel out
each other in the total energy evolution \eqref{Total_Gain}.
The overall energy change is dominated by the perpendicular magnetic gradient
($\mu \vec{v_{E}} \cdot \nabla B$) and curvature
$(\gamma v_{\parallel})^{2} \vec{v_{E}} \cdot [ (\vec{b} \cdot \nabla) \vec{b}
]$, which are proportional to the electron energy due to $\mu$ and
$(v_{\parallel})^{2}$, respectively.
The increase of the electron acceleration with the increasing initial energy
is therefore expected (see the lower-left panel of Fig.\ref{0_0_Trapped}).
(In addition, trapped electrons with larger initial velocities bounce and pass
accelerators more frequently to gain more energies.)

Also due to the combined actions between the magnetic gradient and curvature
in Eq.\eqref{Total_Gain}, the favourable initial pitch
angles (shown in the lower-middle panel of Fig.\ref{0_0_Trapped})
are not 0, $180^{\circ}$ or $90^{\circ}$ which correspond to acceleration
dominated only by magnetic curvatures
($(\gamma v_{\parallel})^{2} \vec{v_{E}} \cdot [ (\vec{b} \cdot \nabla)
\vec{b}]$)
or gradients
($\mu \vec{v_{E}} \cdot \nabla B$), respectively.
No favourable initial pitch angle with 0, $180^{\circ}$ or $90^{\circ}$
indicates that the perpendicular magnetic gradient and curvature acceleration
efficiencies are comparable with each other in this complex
magnetic field structure no matter what resolution of magnetic fields are used.
These favourable initial pitch angles also change with the magnetic field resolution.
The acceleration asymmetry around initial pitch $90^{\circ}$
(the bottom-middle panel of Fig.\ref{0_0_Trapped}) is due to the
non-symmetric acceleration factors around the current sheet center in
($\mu \vec{v_{E}} \cdot \nabla B$) and
$(\gamma v_{\parallel})^{2} \vec{v_{E}} \cdot [ (\vec{b} \cdot \nabla) \vec{b}
]$ by the third dimension of electromagnetic fields in 2.5D symmetric
current sheet geometry.
The acceleration symmetry is weakly broken for the coarsely resolved case.

            \begin{figure}[htbp]
                \centering
                \mbox{
                      \includegraphics[width=0.45\textwidth]{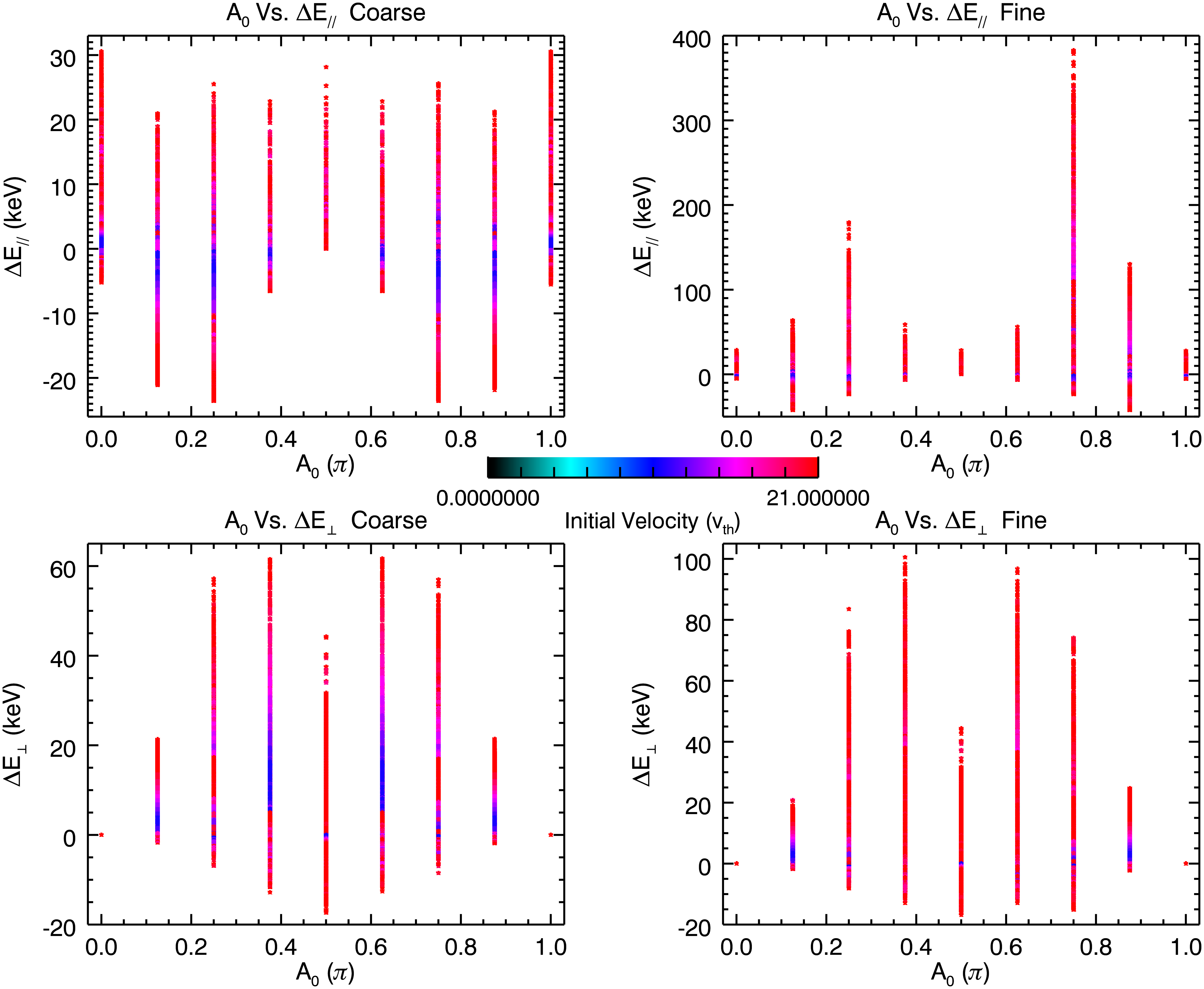}
                     }
                \caption{Parallel ($\Delta E_{\parallel}$ in \textit{top} line) and perpendicular ($\Delta E_{\perp}$ in \textit{bottom})
                         acceleration symmetry about the initial pitch angle $90^{\circ}$.
                         \textit{Left} panels: case of the coarsely resolved magnetic fields.
                         \textit{Right} panels: highly resolved ones.
                         Every trapped electron (shown by one \textit{'$\ast$' point}) is color-coded by its initial velocity.
                         The corresponding color-code is shown in the middle.
                         }
                \label{Symmetry}
                  \end{figure}


Fig.\ref{Symmetry} shows electron acceleration symmetry around the
initial pitch angle $90^{\circ}$ for the parallel and perpendicular energy
gain components.
With the details of the parallel and perpendicular acceleration, one can see
that electron parallel and perpendicular acceleration are not exactly symmetric
around the initial pitch angle $90^{\circ}$ in both the coarsely and finely
resolved magnetic fields.
The symmetry is better preserved in the coarsely resolved fields due to
the smoothing effects (see the bottom-middle panel of Fig.\ref{0_0_Trapped}).

The term $\mu v_{\parallel} (\vec{b} \cdot \nabla B)$ in
Eqs.\eqref{Parallel_Gain} and \eqref{Perpendicular_Gain}, however, is not
influenced by the third dimension of electromagnetic fields, hence it is
symmetric around the current sheet center at the beginning.
In other words, the non-symmetric acceleration in the parallel and
perpendicular direction (see the top and bottom panels
of Fig.\ref{Symmetry}, respectively) are due to non-symmetric
($\mu \vec{v_{E}} \cdot \nabla B$) and
$(\gamma v_{\parallel})^{2} \vec{v_{E}} \cdot [(\vec{b} \cdot
\nabla)\vec{b}]$ around the current sheet center, respectively.
For the coarsely resolved case, the acceleration is dominated by the
perpendicular component, the reverse is true for the finely resolved case.
Meanwhile non-symmetric parallel acceleration
(i.e. non-symmetric $|v_{\parallel}|$) can enhance the
asymmetry in both the parallel and perpendicular acceleration
(see Eqs.\eqref{Parallel_Gain} and \eqref{Perpendicular_Gain}).



            \begin{figure}[htbp]
                \centering
                \mbox{
                      \includegraphics[width=0.45\textwidth]{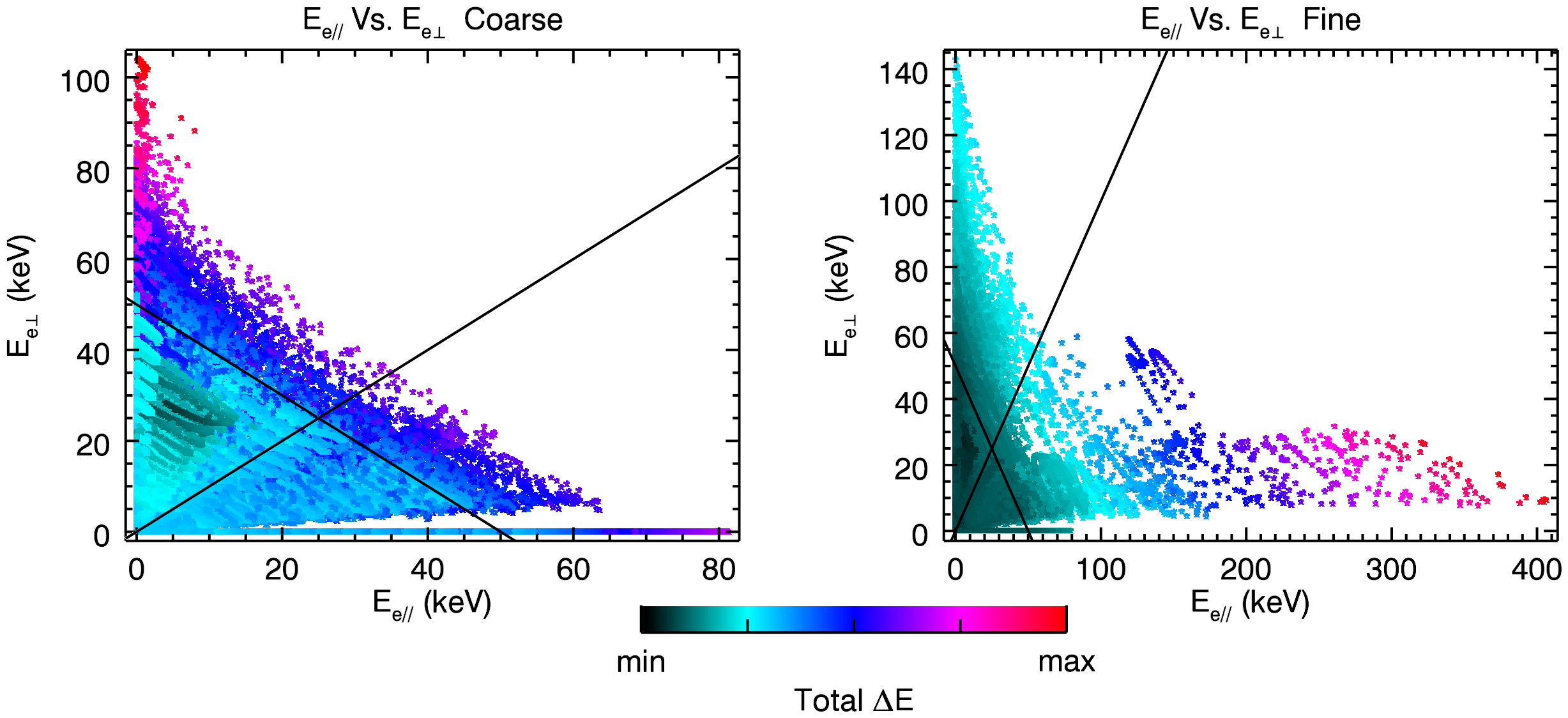}
                     }
                \caption{Comparison between the final parallel ($E_{\parallel}$)
                         and perpendicular ($E_{\perp}$) kinetic energy of trapped electrons in the
                         coarsely (\textit{left} panel) and finely (\textit{right} panel) resolved magnetic fields.
                         Each electron is color-coded by its total kinetic energy change ($\Delta E$).
                         The \textit{black lines} in each panel correspond to '$E_{\perp}=E_{\parallel}$' and
                         '$E=E_{\perp}+E_{\parallel}=50$ $keV$'.
                         }
                \label{Parallel_Total}
                  \end{figure}


Fig.\ref{Parallel_Total} shows the distribution of the trapped electron acceleration
in the ($E_{\parallel}$, $E_{\perp}$) plane.
The distribution of the strongly accelerated
electrons is highly anisotropic. It is dominated by the perpendicular and
parallel kinetic energy components for the coarsely and finely resolved
magnetic fields, respectively.
While for the weakly accelerated electrons, they still roughly keep
their initial isotropic distribution.
For electrons initially moving only along magnetic field lines
(with an initial pitch angle 0 or $180^{\circ}$), they do not have acceleration
in the perpendicular direction.
In this condition, however, the parallel
acceleration is a little stronger in the coarsely resolved magnetic fields
(maximum finale energy 81 $keV$) than that
in the fine case with maximum finale energy 79 $keV$.

\subsubsection{Characteristic trajectories}

\begin{figure}[htbp]
                \centering
                \mbox{
                      \includegraphics[width=0.45\textwidth]{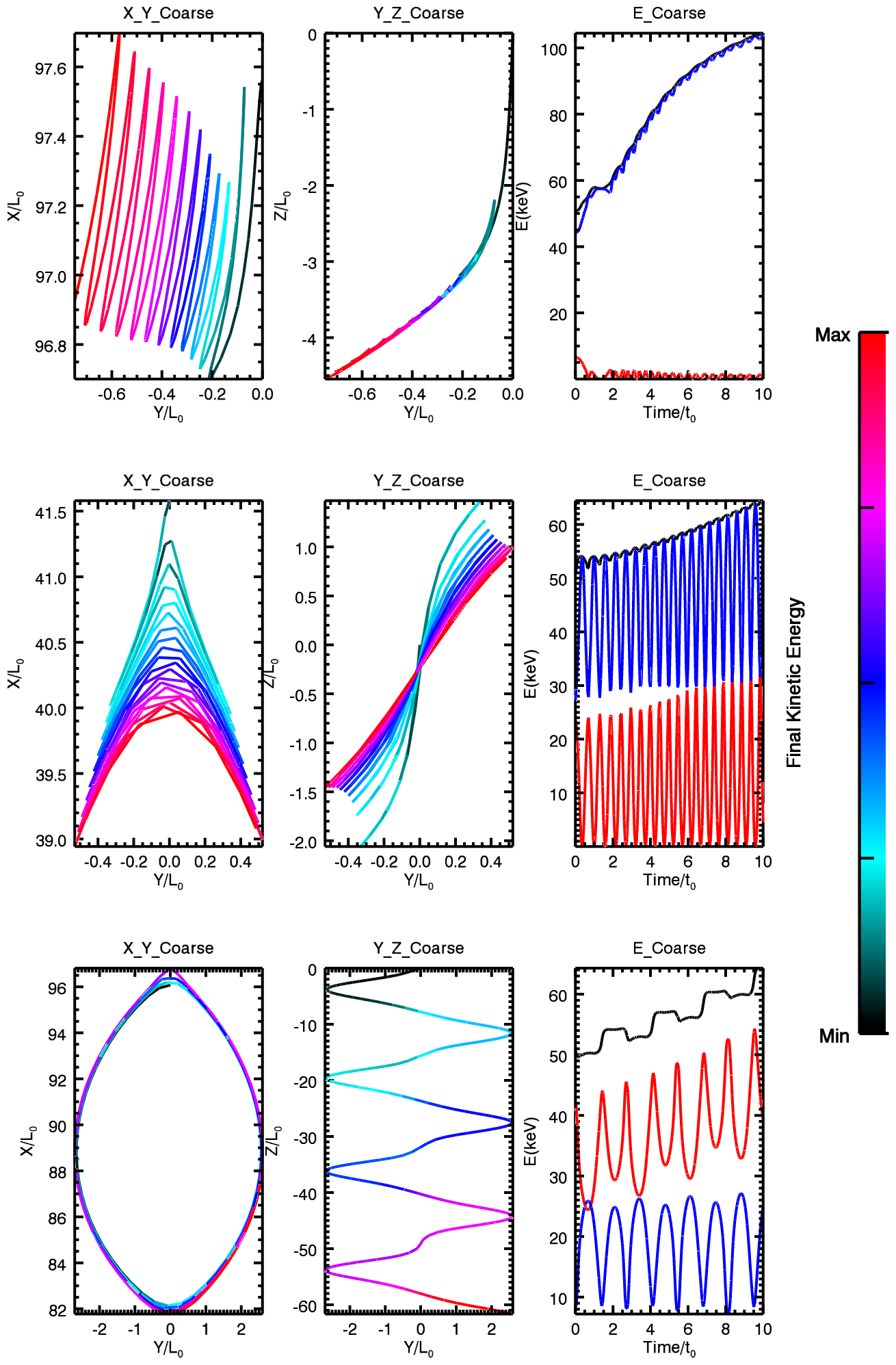}
                     }
                \caption{Trajectory and energy evolution for three characteristic trapped electrons in the coarsely resolved magnetic fields.
                         Panels in the \textit{first and second} column show the electron trajectory in the $xy$ and $yz$ plane, they are
                         color-coded by the total kinetic energy profile.
                         Panels in the \textit{third} column show the electron total kinetic (\textit{black} line), parallel (\textit{red} line)
                         and perpendicular (\textit{blue} line) energy evolution.
                         }
                \label{3Examples_Coarse}
                  \end{figure}

\begin{figure}[htbp]
                \centering
                \mbox{
                      \includegraphics[width=0.45\textwidth]{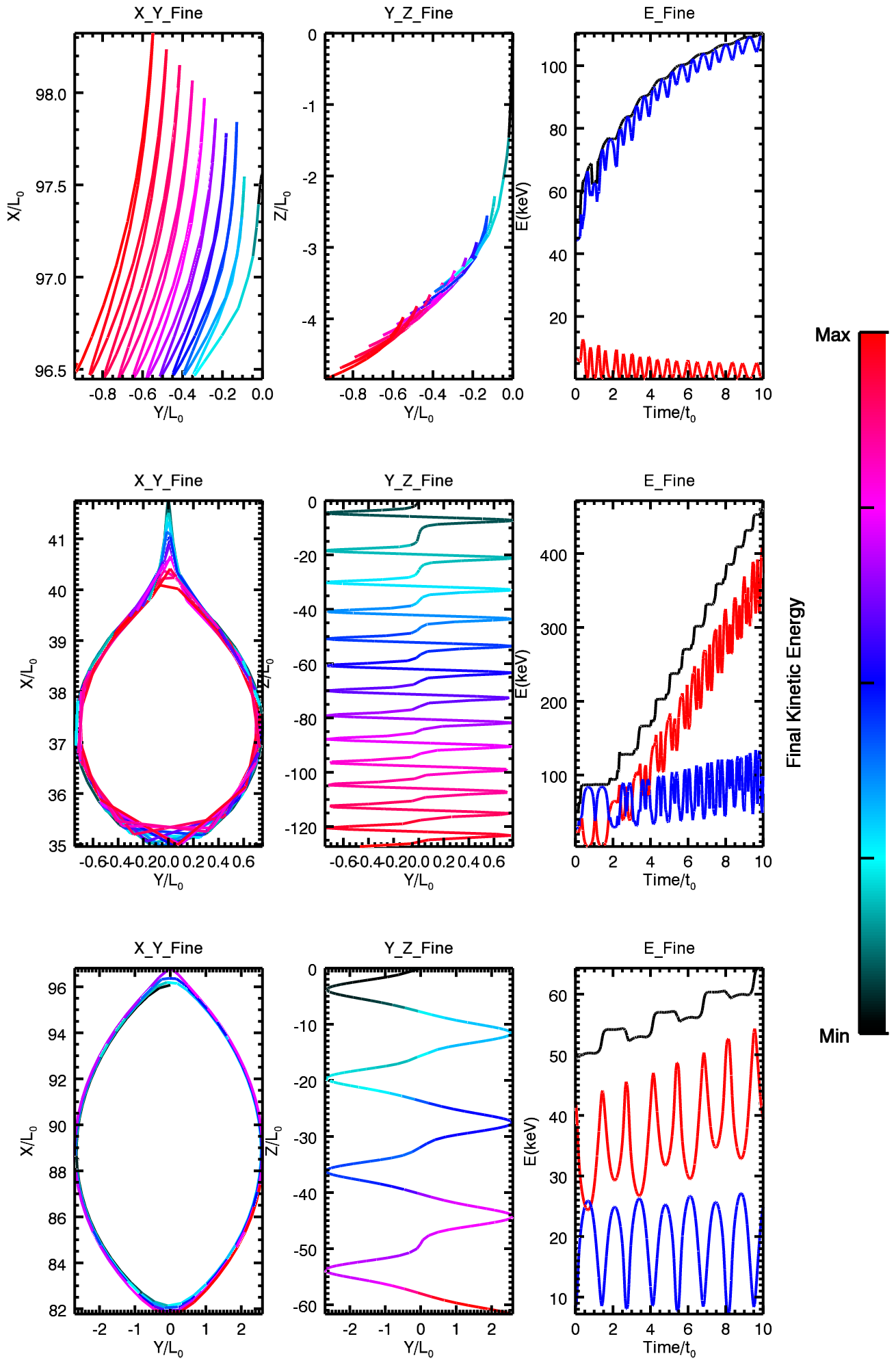}
                     }
                \caption{Trajectory and energy evolution of electrons launched with the same initial conditions as the corresponding one
                         in Fig.\ref{3Examples_Coarse} but in finely resolved magnetic fields.
                         }
                \label{3Examples_Fine}
                  \end{figure}

To better understand the details of the electron acceleration processes,
the first row of Fig.\ref{3Examples_Coarse} and second row of
Fig.\ref{3Examples_Fine} depict the trajectory and energy evolution of
the most energetic electrons in the coarsely
($v_{0}=21$ $v_{th}$, $A_{0}=5/8\pi$, $x_{0}=97.56$ $L_{0}$) and
finely
($v_{0}=21$ $v_{th}$, $A_{0}=3/4\pi$, $x_{0}=41.58$ $L_{0}$)
resolved magnetic fields, respectively, and all panels in
Fig.\ref{3Examples_Fine} have the same initial conditions as the
corresponding panels in Fig.\ref{3Examples_Coarse}.


For the coarsely resolved case, the most efficient acceleration is dominated
by the increase of the perpendicular energy component for the positive
perpendicular magnetic gradient $\mu \vec{v_{E}} \cdot \nabla B$ above 95
$L_{0}$ (see the left panel of Fig.\ref{Acc_Factors}).
Also the slightly decreasing parallel energy is due to the
negative perpendicular curvature
$\vec{v_{E}} \cdot [(\vec{b} \cdot \nabla)\vec{b}]$ there
(see the right panel of Fig.\ref{Acc_Factors}).
With the same initial conditions, the corresponding electron in the finely
resolved magnetic fields gain more energy (see the first row in Fig.\ref{3Examples_Fine}).

For the finely resolved case, the most efficient acceleration happens to the
parallel energy component with slightly increased perpendicular energy when
the electron are trapped in the magnetic island (around $x=37$ $L_{0}$) and
accelerated again and again with its circulating motions by the positive
perpendicular magnetic curvature $\vec{v_{E}} \cdot [(\vec{b} \cdot \nabla)\vec{b}]$
(located at the thin layer in the central current sheet above $x=40$ $L_{0}$,
corresponding to the step-like increased displacement in the $z$ direction,
see the final kinetic energy color-coding $yz$-trajectory projection in the
second row of Fig.\ref{3Examples_Fine})
and
gradient $\mu \vec{v_{E}} \cdot \nabla B$ (located around $x=40$ $L_{0}$).
As expected, in the coarsely refined magnetic fields without the smaller-scale
magnetic field structures, trajectory and acceleration of the corresponding
electron are totally changed. This electron is mirror-trapped and cannot
circulate in the magnetic
island and move systematically in the $z$ direction.

In the right panel of Fig.\ref{Acc_Factors}, one can see the largest
perpendicular curvature acceleration region is located at around $x=85$ $L_{0}$,
however, the strongest energetic electron is not launched there.
That is due to the cancellation between the perpendicular curvature acceleration
and deceleration when electron circulates in the magnetic island at
$x \sim 90$  $L_{0}$.
Electrons in the third row of Fig.\ref{3Examples_Coarse} and
Fig.\ref{3Examples_Fine} are chosen to reveal the acceleration characteristics
around $x=90$ $L_{0}$.
These two electrons also have the same initial conditions:
$v_{0}=21$ $v_{th}$, $A_{0}=7/8\pi$ and $x_{0}=96.075$ $L_{0}$.
Their color-coded $yz$-trajectory projection in the third
row of Fig.\ref{3Examples_Coarse} and Fig.\ref{3Examples_Fine} prove the above
discussion and their energy gains are mainly due to the parallel acceleration
by the perpendicular magnetic curvature.
Furthermore these two electrons have the same trajectory and energy
evolution, i.e., there is no influence coming from the magnetic field
resolution, since no refined
smaller-scale magnetic structures are found along their trajectories
between $x=81 - 97$ $L_{0}$.

Energy oscillation between parallel and perpendicular energies in each
characteristic electron energy evolution profile (the last column of
Fig.\ref{3Examples_Coarse} and Fig.\ref{3Examples_Fine} ) is due to the
parallel magnetic gradient $v_{\parallel} (\vec{b} \cdot \nabla B)$ in
Eqs.\eqref{Parallel_Gain} and \eqref{Perpendicular_Gain} when electron passes
the positive and negative
parallel magnetic gradient regions in turn
or
electron is mirrored with alternate parallel velocity in the parallel and
anti-parallel direction.
Each condition can be found in Fig.\ref{3Examples_Coarse} and
Fig.\ref{3Examples_Fine}.
While the magnitude of this oscillations is
due to magnitude of the parallel magnetic gradients $\vec{b} \cdot \nabla B$
along electron trajectory (see the middle panel of Fig.\ref{Acc_Factors}).

\subsubsection{Comparison with Observations}


More than $60\%$ of trapped electrons are accelerated
($\Delta E>0$) and more than $50\%$ of them have kinetic energies
larger than 10 $keV$.
These energetic electrons can produce HXRs by Bremsstrahlung (note that
the HXR range is $10 - 400$ $keV$).
In the framework of the thin target model (\citealp{Brown1971}) and using the
Bethe-Heitler formula for the Bremsstrahlung cross section
(\citealp{1934RSPSA.146...83B, Brown1971}, see
details in Sec.\ref{Method2}), we derive the HXR spectrum of these energetic
trapped electrons in order to compare our results with solar flare HXR
observations.

Since the initial electron distribution function in the solar atmosphere is not
known, we consider three different initial distribution functions as:
\begin{align}
f(E_{0}, A_{0}, \vec{r_{0}}, t_{0}) \propto
\left \{
             \begin{array}{cc}
                   E_{0}^{0} \texttt{ (or Constant)}                                  \\
                   E_{0}^{-3}                                \\
                   \texttt{Maxwell-Boltzmann}
             \end{array}
\right.
\label{Initial_Distribution_Function}
\end{align}
%

                  \begin{figure}[htbp]
                  \centering
                      \mbox{
                            \includegraphics[width=0.45\textwidth]{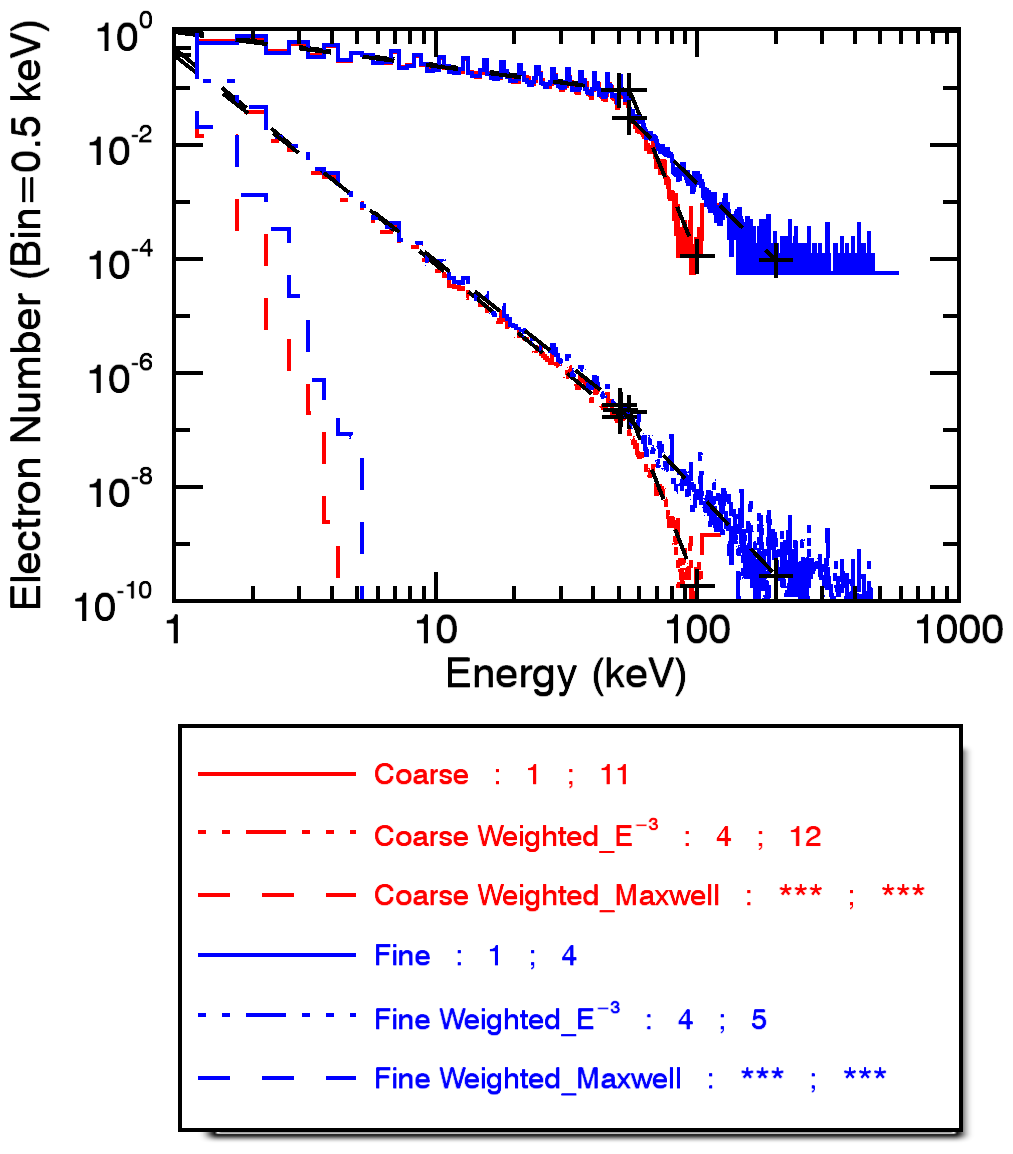}
                           }
                           \\
                      \mbox{
                            \includegraphics[width=0.45\textwidth]{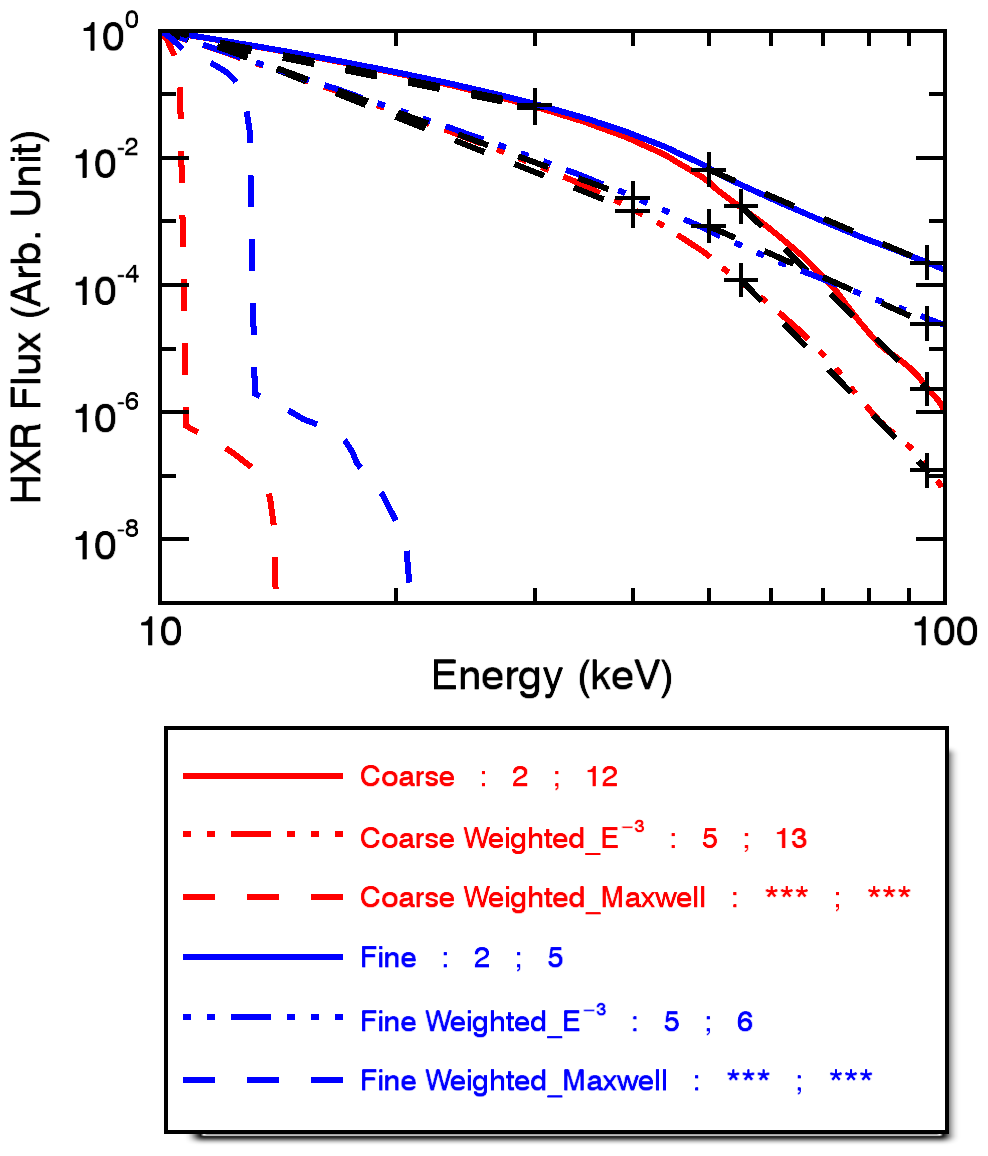}
                           }
                      \caption{Electron (\textit{top}) and HXR (\textit{bottom}) spectra of energetic trapped electrons in
                               the coarsely (\textit{red} lines) and finely (\textit{blue} lines) resolved magnetic fields with
                               three different initial distribution functions - constant (\textit{solid} lines),
                               power-law $-3$ (\textit{dash-dot} lines) and Maxwellian at $10^6$ $K$ (\textit{dashed} lines).
                               The spectral indices are for the ranges marked with the black dashed lines embraced by two plus signs at two ends
                               and their values are shown under each panel:
                               the first one for electron and photon energies below 50 $keV$ and the second one for the energies
                               between $50$ $keV$ and $100$ $keV$.
                             }
                      \label{0_HXR}
                  \end{figure}

The resulting electron and HXR spectra and spectral indices (below and above
50 $keV$) after acceleration ($t=10$ $t_{0}$)
are depicted in the top and bottom two panels of Fig.\ref{0_HXR}, respectively.
In approximation, the relationship between the electron ($\gamma_{e}$) and
corresponding HXR ($\gamma_{HXR}$) spectral indices agree well with the
relationship $\gamma_{HXR}=\gamma_{e}+1$ in the thin target model.
The influence from the ambient plasma number density $n_{\vec{r}}$
(Eq.\eqref{HXREquation}) is very small due to it small normalized range
$[0.3 - 1.8]$ in both differently resolved magnetic fields, while the HXR
flux at 100 keV differs by more than 4 orders of magnitude for these two cases
(see the bottom panel of
Fig.\ref{0_HXR}).
Note that the HXR spectral indices, calculated from an initial
Maxwell-Boltzmann distribution function for $T=10^6$ K, are too large to match any
observed HXR spectrum.
For the cases with a power-law distribution initially
(e.g., \citealp{Karlicky&Barta2006}), one may treat the electron acceleration
as a diffusion in 2D energy space. Since the diffusion coefficient is
approximately proportional to the energy, this explains the difference by
one of the spectral indexes of the injected and accelerated electrons below
the maximum injection energy of $\sim 50$ $keV$.
Above 50 $keV$, the acceleration in the finely resolved magnetic field is
much more efficient than that in the coarsely resolved magnetic field,
we have a harder spectrum ($\sim 6$) for the finely resolved case.
The corresponding HXR spectral indexes are consistent with the observed
values for small flares (whose HXR spectral indices can be as soft as
$\geq 7$, see \citealp{Aschwanden2002}).

                  \begin{figure}[htbp]
                  \centering
                      \mbox{
                            \includegraphics[width=0.45\textwidth]{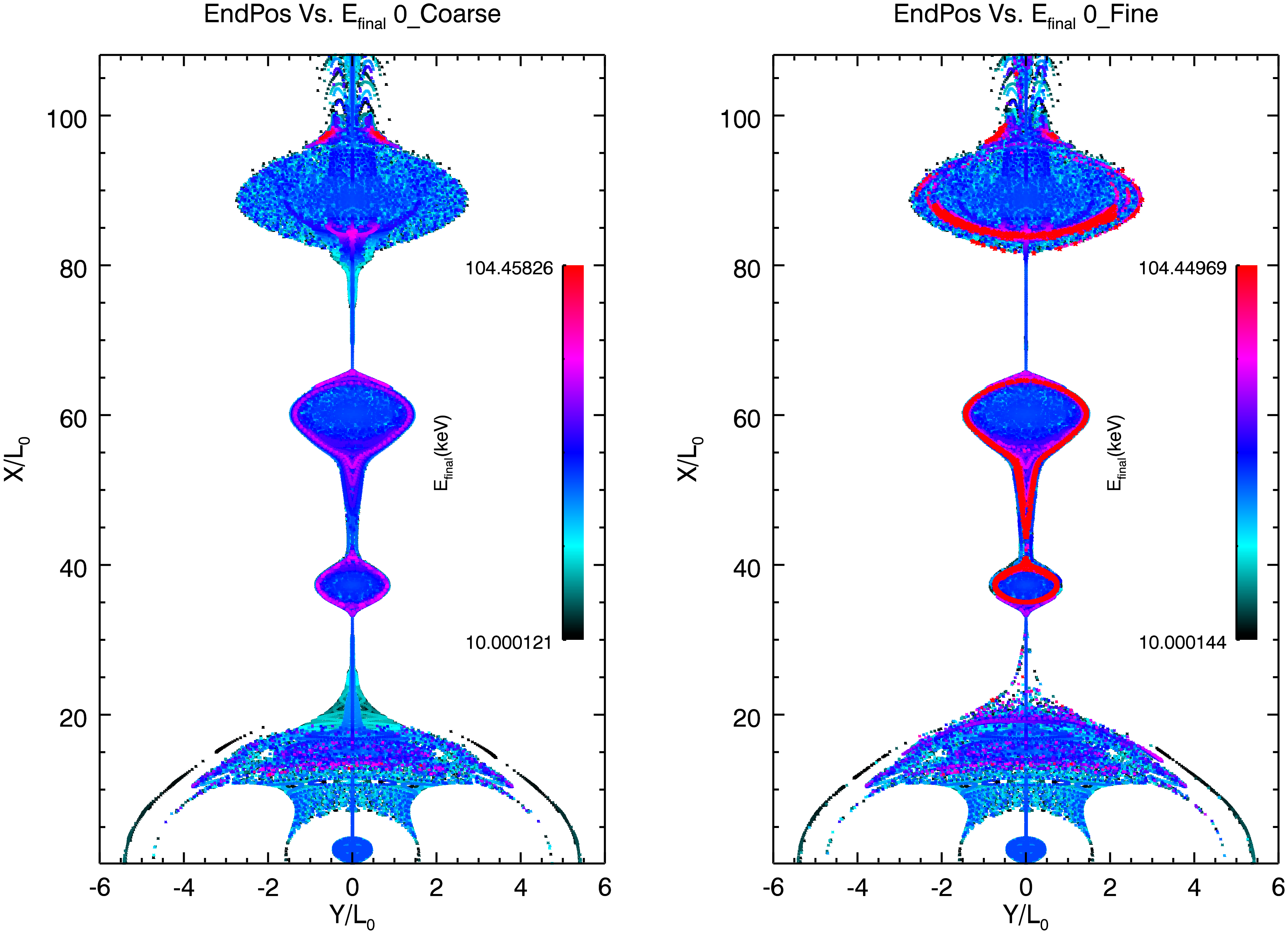}
                           }
                      \caption{Final locations of trapped electrons with final kinetic energies $> 10$ $keV$ at $t=10t_{0}$.
                               They are color-coded by their final kinetic energies.
                               \textit{Left} and \textit{right} panels correspond to the
                               coarsely and finely resolved magnetic fields.
                               Note that the electrons with final kinetic energy $>105keV$
                               are shown only by \textit{red asterisk} points in the better resolved magnetic fields.
                             }
                      \label{EP_Trapped}
                  \end{figure}

Besides the HXR spectra, fine structures (bright spots) along the
current sheets trailing CMEs or eruptive filaments were observed
(e.g., by \citealp{Ciaravellaetal2002,
Koetal2003, Savage2010}).
These bright spots should come from energetic trapped electrons.
Fig.\ref{EP_Trapped} shows the final locations of these
trapped electrons with final kinetic energies $>10 keV$ which
will brighten the magnetic island in the current sheet that may be associated
with the observed hot spots.
Furthermore depending on the evolution of these magnetic islands, the bright
spot located at $x=90$ $L_{0}$ moves upwards away from the sun while others
fall back to the sun.
This evolution agrees well with the observed
upward (\citealp{Ciaravellaetal2002, Koetal2003, Savage2010}) and
downward (\citealp{Koetal2003, Savage2010}) moving bright spots in CME-trailing
current sheets.

          \subsection{Precipitating Electrons}
          \label{0_Fled}

\begin{figure}[htbp]
                \centering
                \mbox{
                       \includegraphics[width=0.25\textwidth]{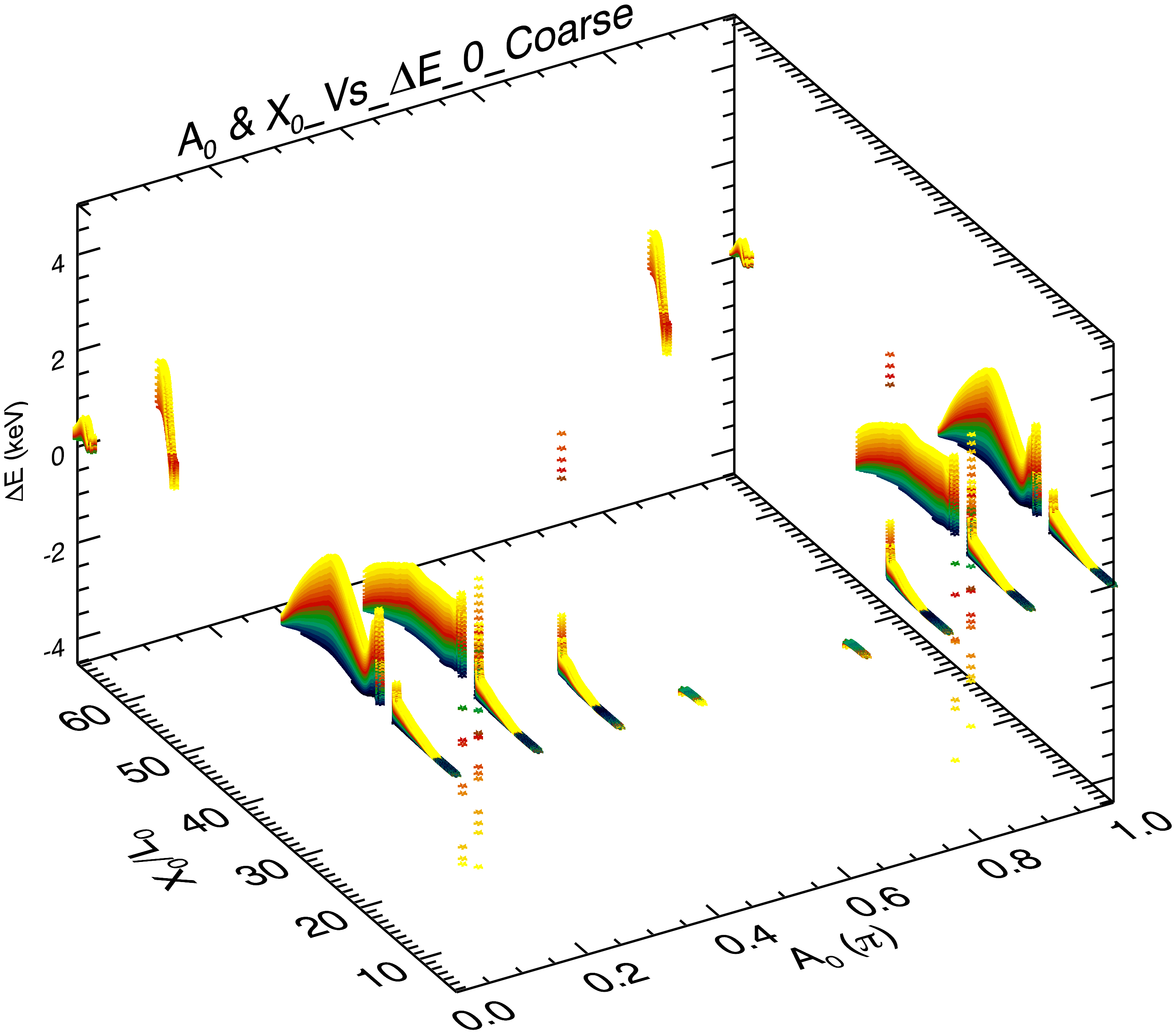}    
                       \includegraphics[width=0.25\textwidth]{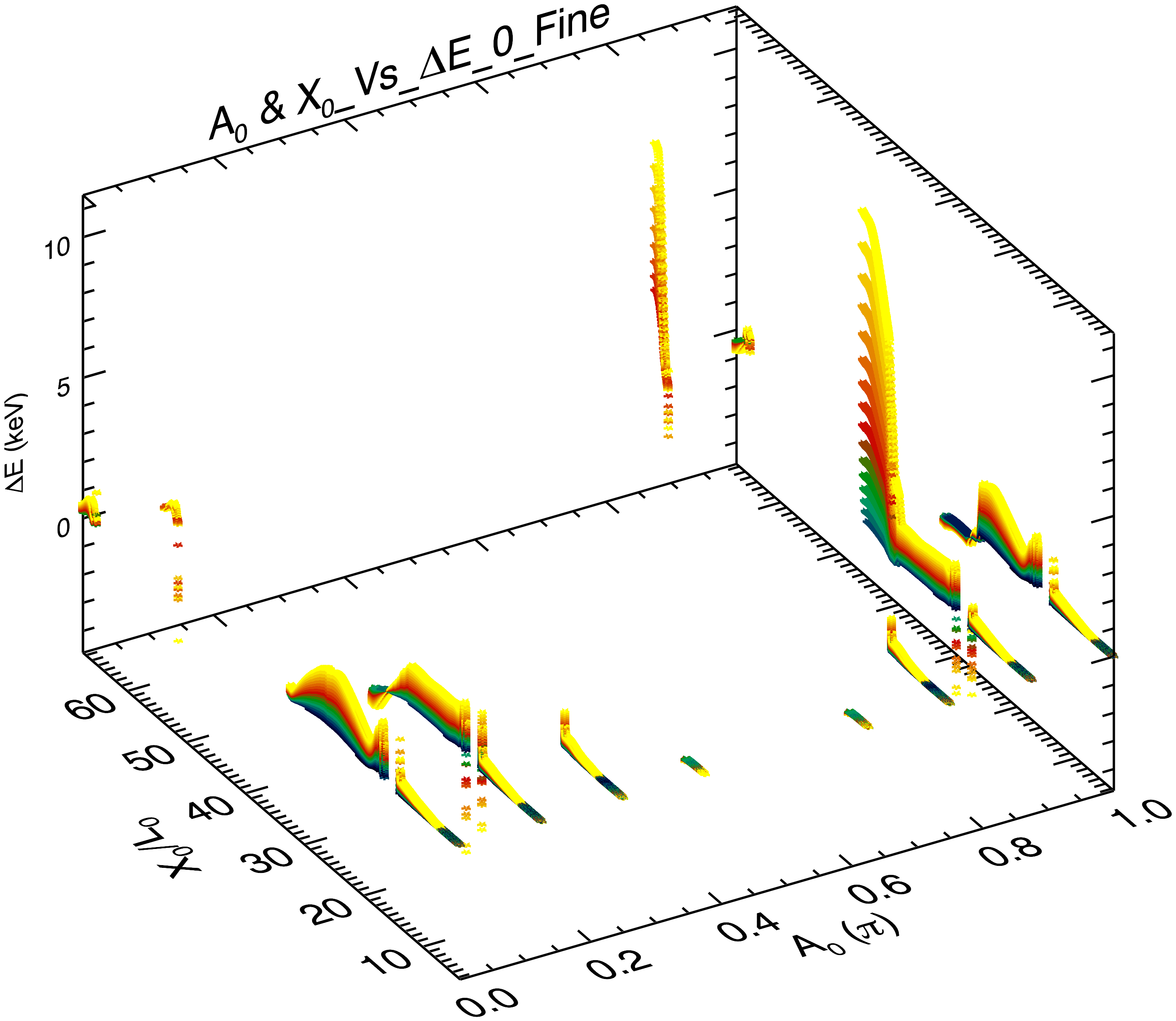}
                     }
                    \\
                \mbox{
                       \includegraphics[width=0.50\textwidth]{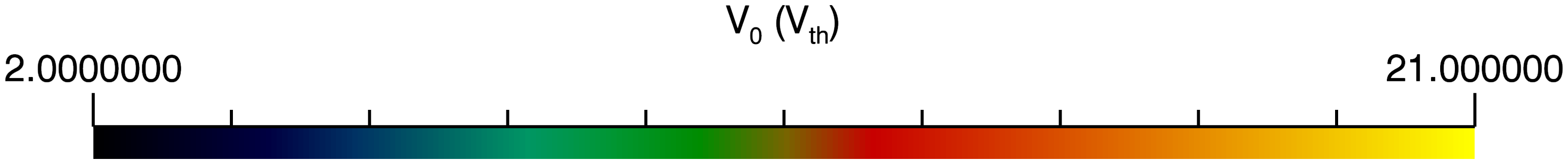}
                     }
                     \\
                \mbox{
                       \includegraphics[width=0.50\textwidth]{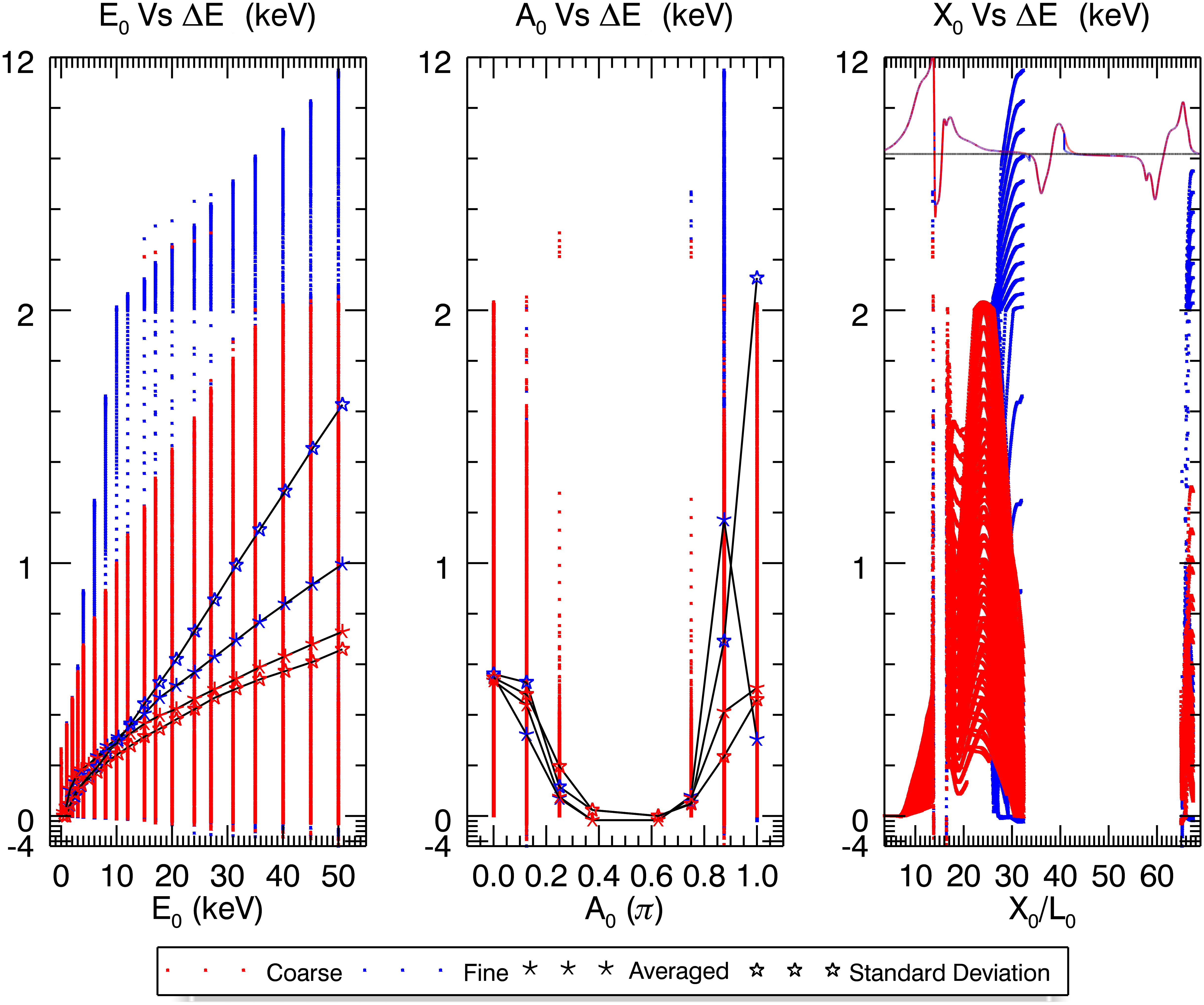}
                     }
                \caption{Same as Fig.\ref{0_0_Trapped} but for precipitating electrons.
                         Here three different scales in the y-axis are also used for $\Delta E <0$, $0<\Delta E<2$ $keV$ and
                         $\Delta E > 2$ $keV$.
                        }
                \label{0_0_Fled}
            \end{figure}

A second observable feature which can be derived from our calculations are the
emission produced by the energetic precipitating electrons. They can
precipitate to the solar
chromosphere and be related to the observed footpoint HXR signatures there.
We first study the acceleration dependence of
precipitating electrons on the initial conditions (velocity,
pitch angle and position). The results are shown in Fig.\ref{0_0_Fled}.
\subsubsection{Initial condition dependence}

Similar to electrons trapped in the current sheet,
acceleration of precipitating electrons also strongly depends on their
initial (velocity, pitch angle and position) conditions.
The acceleration efficiency increases with the increase of the energy and
the overall acceleration is more efficient in finely resolved magnetic field.
However the acceleration is much less efficient than those trapped electrons.
The maximum energy gain is only a few $keV$ and about $10$ $keV$ for the coarsely
and finely resolved magnetic fields respectively.
The dependence of the energy gain on the initial pitch angle and position
show that only electrons in a few channels can escape from the acceleration
site and injected into the chromosphere.
As expected, electrons moving along
magnetic field line are more likely to escape than those with a pitch angle
close to $90^{\circ}$.
However, only a small portion ($< 12\%$) of electrons
can precipitate into the chromosphere.

            \begin{figure}[htbp]
                \centering
                \mbox{
                      \includegraphics[width=0.45\textwidth]{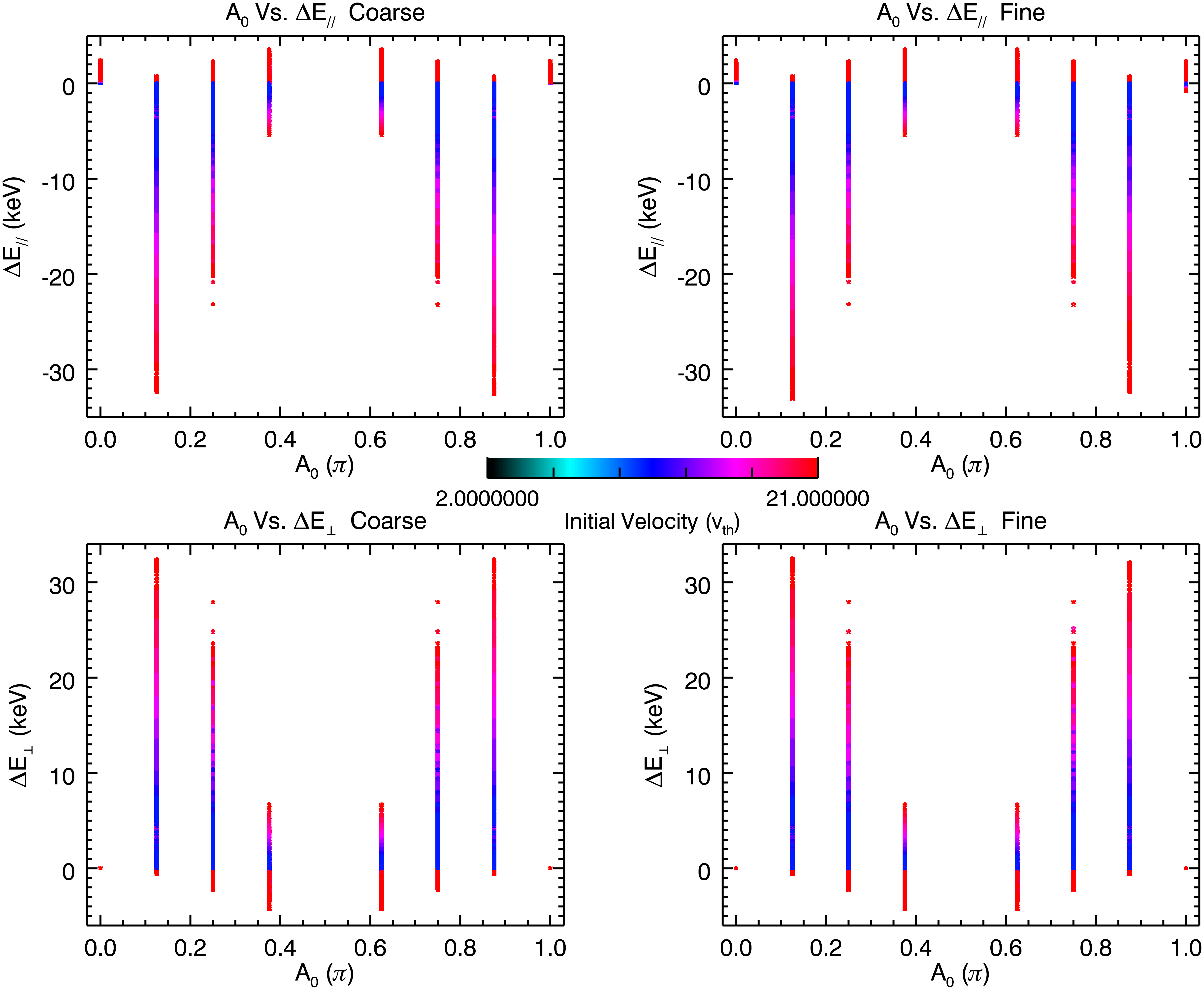}
                     }
                \caption{Same as Fig.\ref{Symmetry} but for precipitating electrons.
                         }
                \label{Symmetry_Precipitating}
                  \end{figure}

The bottom right panel of Fig.\ref{0_0_Fled} depicts that a large portion of
precipitating electrons start near to X-points.
But no electron escapes from the X-points near $x=42$ $L_0$ and $x=97$ $L_0$.
Since the magnetic islands below this two X-points is not symmetric about
their center - O-points: their upper parts are smaller than the lower parts (see
the whole $B_{y}$ plots in the bottom right panel of Fig.\ref{0_0_Trapped}).
Current sheet center launched electrons are easily reflected or trapped by this
geography (see the characteristic trajectories of trapped electrons launched
near $x=42$ $L_0$ in the middle-line of Figs.\ref{3Examples_Coarse} and
\ref{3Examples_Fine}).

\subsubsection{Acceleration properties}

Although there are the same reasons for the asymmetric acceleration around the
initial pitch angle $90^{0}$ in the finely resolved magnetic
fields between trapped and precipitating electrons, acceleration asymmetry
of precipitating electrons is much weaker than that of trapped electrons
(comparing Fig.\ref{Symmetry} with Fig.\ref{Symmetry_Precipitating}).

\begin{figure}[htbp]
                \centering
                \mbox{
                      \includegraphics[width=0.45\textwidth]{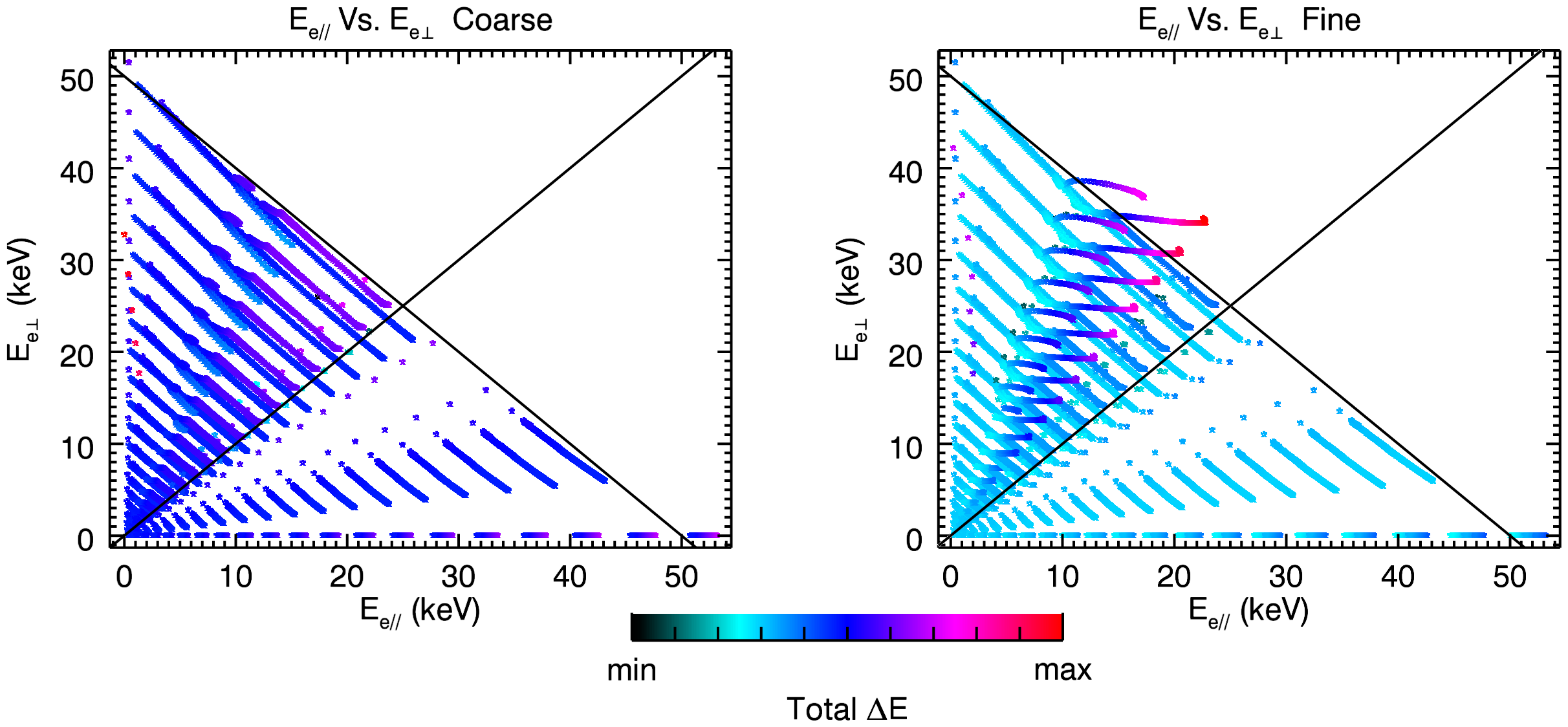}
                     }
                \caption{Same as Fig.\ref{Parallel_Total}, but for
                         precipitating electrons. }
                \label{Parallel_Fled}
            \end{figure}

The total acceleration of precipitating electrons also is much
weaker than that of trapped electrons,  see the '$E=50$ $keV$' parts of
Fig.\ref{Parallel_Total} and Fig.\ref{Parallel_Fled}, especially the coarse
case for precipitating electrons in Fig.\ref{Parallel_Fled}.
The final kinetic energy $E_{e}$ of
the most energetic precipitating electrons is a little more than 50 and 60
$keV$ in the coarsely and finely resolved magnetic fields, respectively,
i.e., all precipitating electrons have final kinetic energies $E_{e}
<100$ $keV$.
Different from trapped electrons in Fig.\ref{Parallel_Total}, here most
precipitating electrons still keep their initial energies shown as stripes
parallel to '$E=50$ $keV$'

Acceleration difference between trapped and precipitating electrons
is mainly contributed by the acceleration in the parallel direction.
Precipitating electrons have stronger deceleration than acceleration in
the parallel direction (see the top panels of Fig.\ref{Symmetry} and
Fig.\ref{Symmetry_Precipitating}), i.e., the acceleration of precipitating
electrons are mainly coming from the perpendicular direction independent on
the magnetic field resolution (see also Fig.\ref{Parallel_Fled}).
For a stronger parallel acceleration, electron should stay longer
around the current sheet center where has larger magnetic curvatures than
other places, while precipitating electrons are ejected out of the current
sheet before they can reach higher energies.

On the whole the acceleration differences of precipitating electrons between
in the coarsely and finely resolved magnetic fields are not so many as that
of trapped electrons, since precipitating electrons spend most of their
time far away from the central current sheet and the better resolved magnetic
structures are located only near the current sheet center.

\subsubsection{characteristic trajectory of precipitating electron}

            \begin{figure}[htbp]
                \centering
                \mbox{
                      \includegraphics[width=0.45\textwidth]{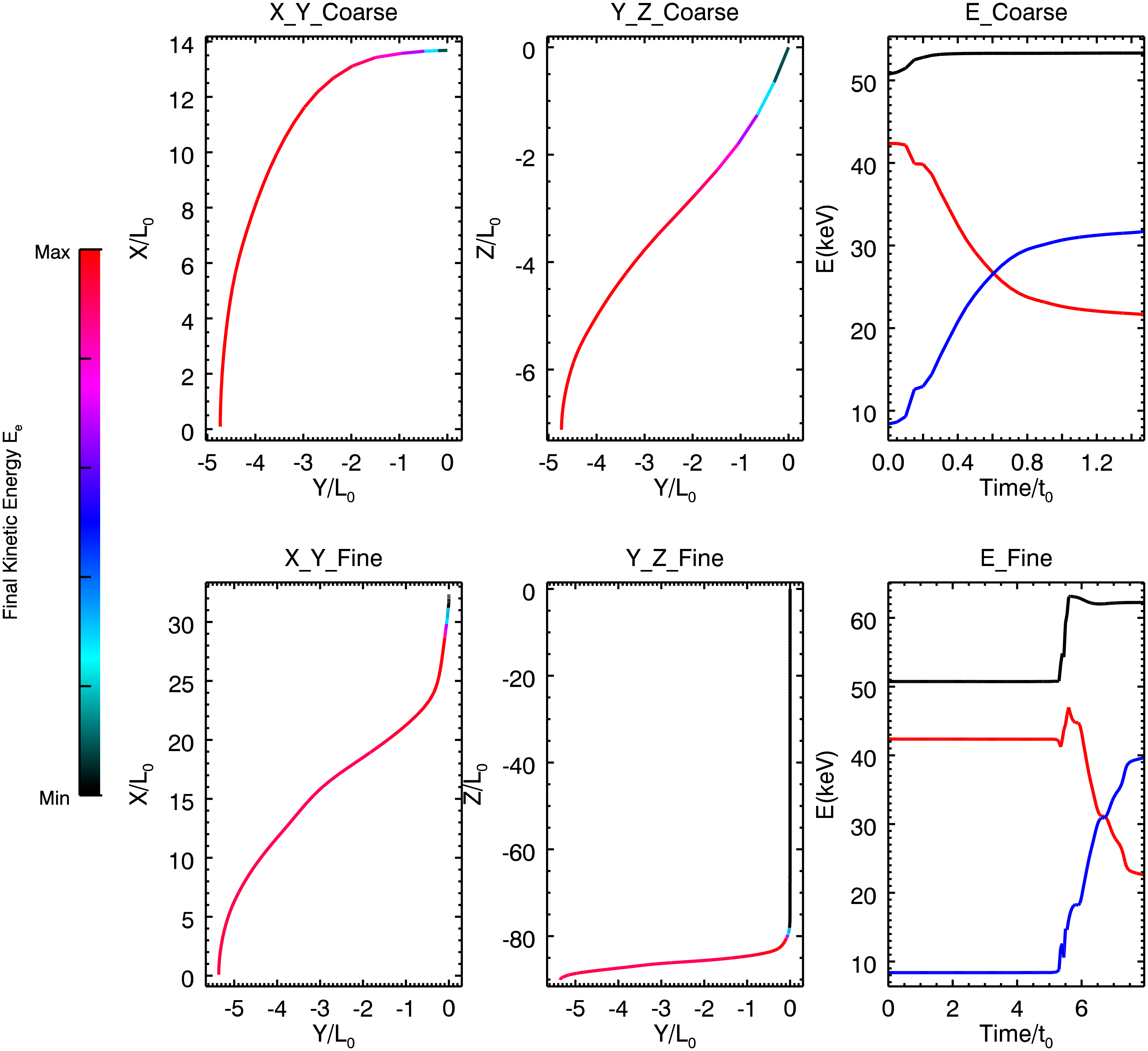}
                     }
                \caption{Trajectory projection and energy evolution of the most energetic precipitating electron with initial velocity
                         $v_{0}=21$ $v_{th}$ and pitch angle $157.5^{\circ}$ in the
                         coarsely (\textit{left} panel) and finely (\textit{right} panel) resolved magnetic fields.
                         Trajectories are color-coded according to the local electron
                         total kinetic energy.
                        }
                \label{Fled_Test}
           \end{figure}

Top and bottom lines of Fig.\ref{Fled_Test} show the trajectory and energy
evolution of the most energetic precipitating electron in the coarsely and
finely resolved magnetic fields, respectively.
Their $xy$ and $yz$ trajectory projections (along the magnetic field lines only)
prove their strongly magnetized condition.
Also the energy profiles in the last column of Fig.\ref{Fled_Test} depict the
different acceleration properties in the coarsely and finely resolved
magnetic fields: their acceleration sites are still in the current sheet
center by the perpendicular magnetic curvatures
$\vec{v_{E}} \cdot [(\vec{b} \cdot \nabla)\vec{b}]$
and gradients $\vec{v_{E}} \cdot \nabla B$.
After they leave there is no acceleration any more, there parallel
magnetic gradient $v_{\parallel} (\vec{b} \cdot \nabla B)$ is stronger than
the other two terms (see Fig.\ref{Acc_Factors}).

Also because of the single sign of the parallel magnetic gradients and
direction of parallel velocity along precipitating electron trajectory,
precipitating electrons do not have frequent energy oscillation as that of the
characteristic trapped electrons in
Fig.\ref{3Examples_Coarse} and Fig.\ref{3Examples_Fine}.

\subsubsection{Comparison with UV and EUV observations}
            \begin{figure}[htbp]
            \centering
                 \mbox{
                      \includegraphics[width=0.45\textwidth]{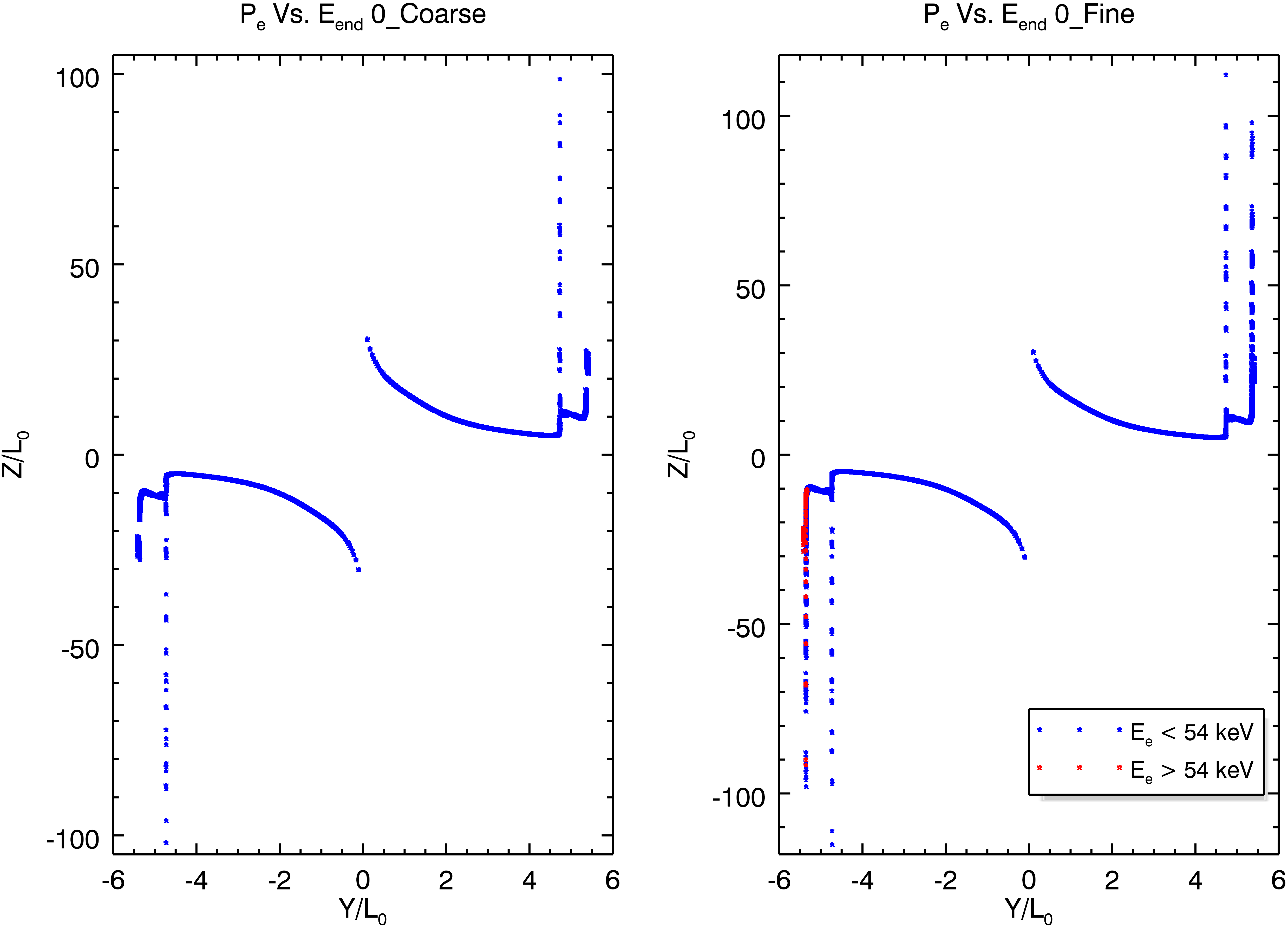}
                      }
                      \\
                 \mbox{
                      \includegraphics[width=0.45\textwidth]{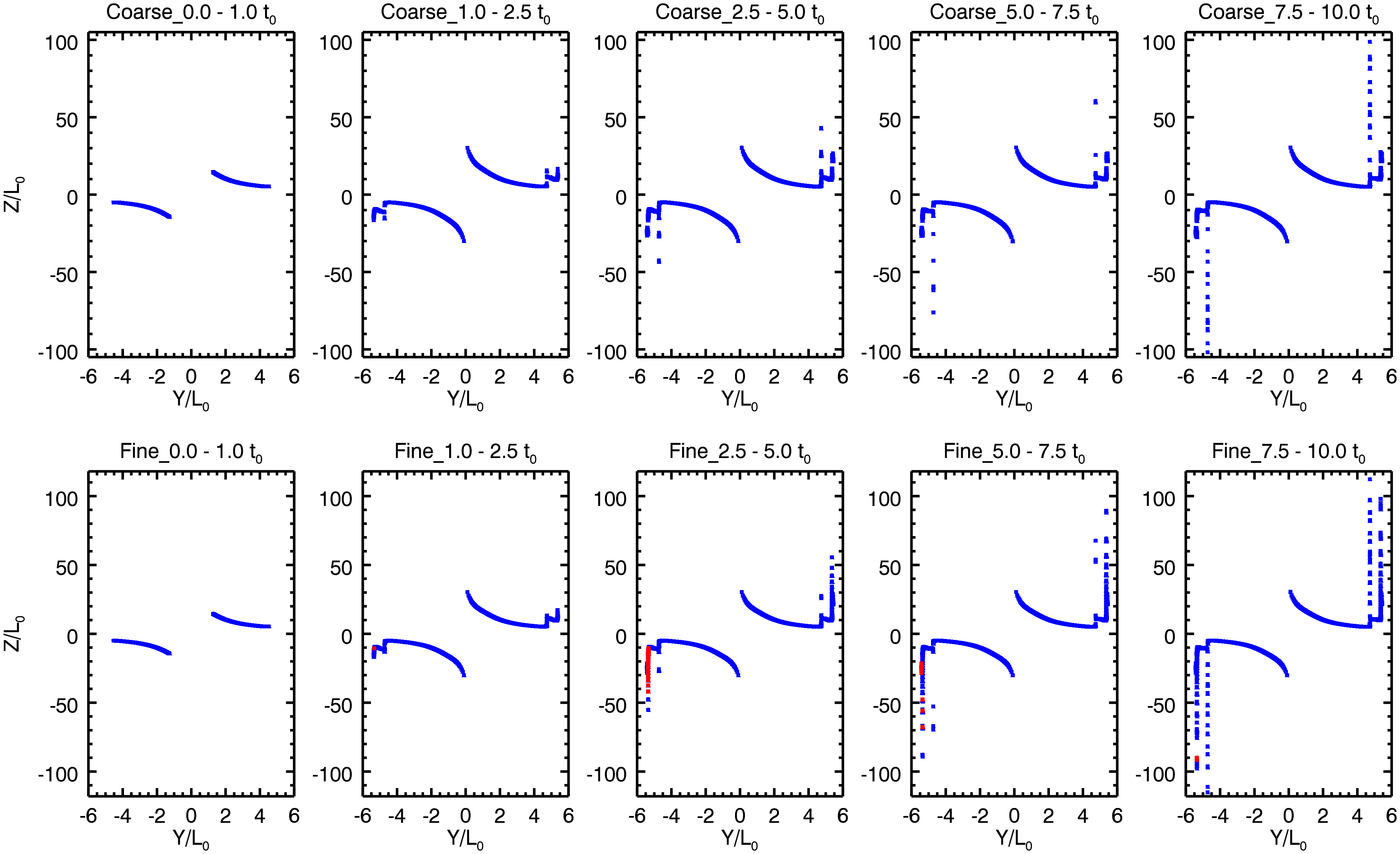}
                      }
                \caption{Chromosphere locations of precipitating electrons at t=10 $t_{0}$ (\textit{top} line) and their evolution
                        (\textit{bottom two} lines), color-coded by their final kinetic energies
                        ($E_{e}$ - \textit{Blue $'\ast'$} for $E_{e} < 54$ $keV$ and \textit{red $'\ast'$} for $E_{e} > 54$ $keV$)
                        separately in the coarsely and finely resolved magnetic islands.
                         }
                \label{0_EndPosition}
            \end{figure}

The low energies of the precipitating electrons in the convective electric
fields  can not cause HXR emissions but ribbons of UV and
EUV brightening (\citealp{Fletcheretal2011}).
Fig.\ref{0_EndPosition} depicts the spatial distribution of the electrons
precipitating to the chromosphere at the end of calculation
(t=10 $t_{0}$, top panels) and
their evolution with time (panels in the last two lines).
As the figure shows the ribbons exhibit a anti-symmetric geometry
around the PIL.

This two ribbons are related to the initial pitch angles of
precipitating electrons:
electron with an initial pitch angles $ > 90^{\circ}$ ($< 90^{\circ}$)
precipitates into one (the other) branch.
This kind of initial pitch angle dependence is attributed to the weak parallel
accelerations by the perpendicular magnetic curvature
$(\gamma v_{\parallel})^{2} \vec{v_{E}} \cdot [(\vec{b} \cdot \nabla)\vec{b}]$
(Eq.\eqref{Parallel_Gain}) which cannot
accelerate electron into the direction anti-parallel to its initial
velocity.
Because of the non-symmetric acceleration around the initial pitch angle
$90^{\circ}$ of precipitating electrons in the finely resolved magnetic fields,
more efficiently accelerated precipitating electrons are with initial
pitch angles 0.875$\pi$ ($>90^{\circ}$).
Hence more accelerated ($E_{e} > 54$ $keV$) precipitating electrons are only
located at one branch of the ribbon geometry with finely resolved
smaller-scale magnetic structures (the top-right panel of Fig.\ref{0_EndPosition}).
Some of the observed asymmetry between two footpoints therefore may be
attributed to the acceleration process.
While the chromosphere energy distribution of precipitating energized electrons,
accelerated by coarsely resolved magnetic structures, are more anti-symmetric
with respect to the PIL (the top-left panel in Fig.\ref{0_EndPosition}).

\begin{figure}[htbp]
            \centering
                 \mbox{
                      \includegraphics[width=0.45\textwidth]{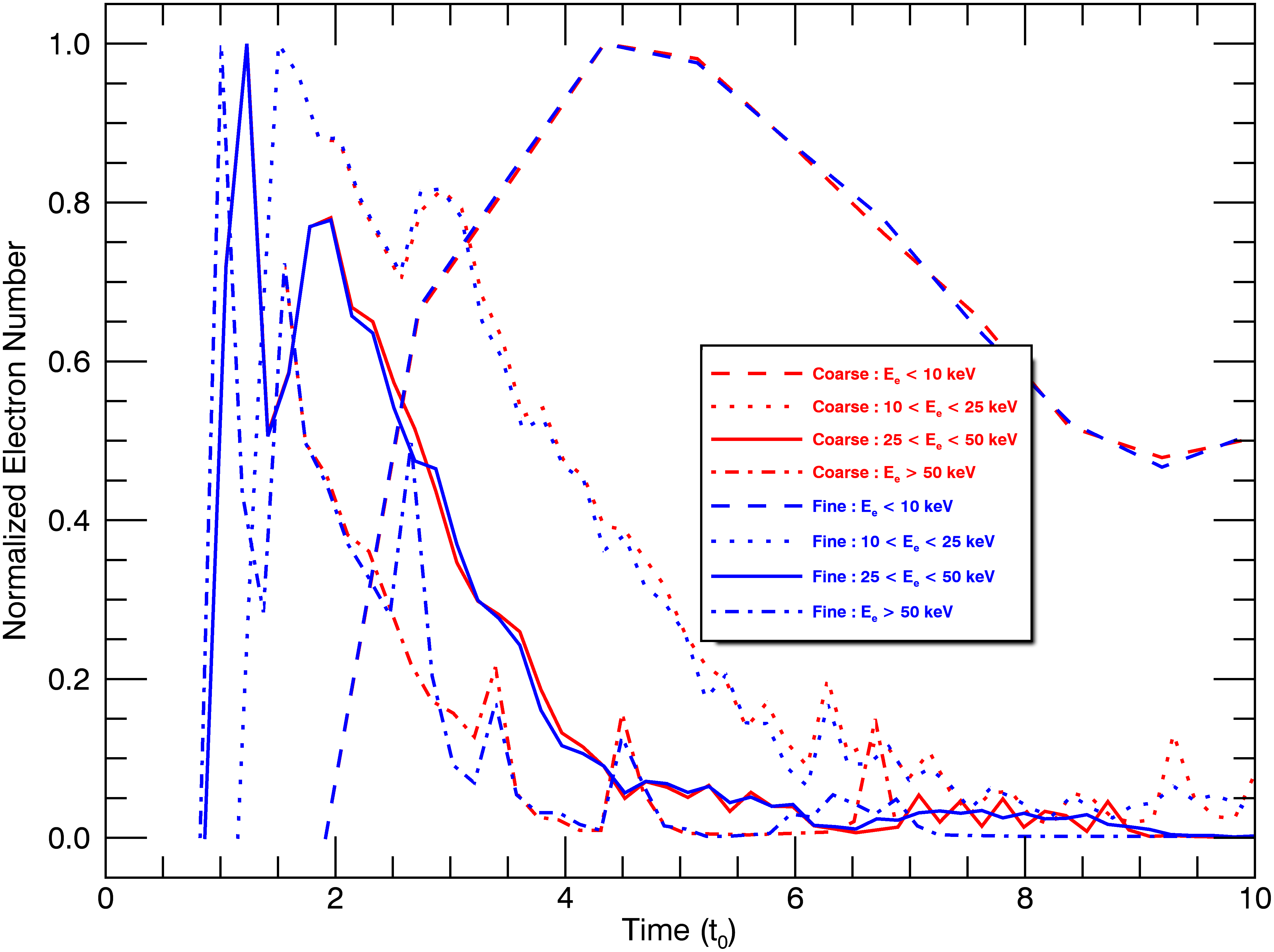}
                      }
                \caption{Lightcurve of precipitating electrons for four energy ranges:
                         $E_{e} < 10 $ $keV$ - \textit{dashed} lines
                         $10< E_{e} < 25 $ $keV$ - \textit{dotted} lines
                         $25< E_{e} < 50 $ $keV$ - \textit{solid} lines
                         $E_{e} > 50 $ $keV$ - \textit{dash-dot} lines for the acceleration in the coarsely
                         (\textit{red} lines) and finely (\textit{blue} lines) resolved magnetic fields.
                         }
                \label{Electron_Light}
            \end{figure}

Also with the chromospheric location evolution of the precipitating electrons
(panels in the last two lines of Fig.\ref{0_EndPosition}),
one can find their locations along the chromospheric ribbons
depend on their initial positions also: electrons started closer to the sun
surface precipitate closer to the PIL , earlier in the
chromosphere and have shorter displacements along z-axis (or PIL).
At the same initial position, electrons with larger
initial energies correspond to larger final kinetic energies and parallel
velocities which lead electron to reach chromosphere earlier (see
Fig.\ref{Electron_Light}).

Fig.\ref{Electron_Light} shows that fluxes of electrons with higher energies
evolve faster and reach peaks earlier than those of lower energy electrons.
The time scale of the flux peak of precipitating
electrons with final kinetic energies $> 50 keV$ indicates precipitating
electrons are accelerated less than 1.0 $t_{0}$ $< 0.1$ $s$.
The refined magnetic field structures are mainly located above $x=25$ $L_{0}$, hence
at the beginning (before 1.5 $t_{0}$) there is no acceleration difference
between the coarsely and finely resolved magnetic fields for precipitating
electrons.

%
%
\section{Conclusions and Discussion }
\label{Discussion}

\subsection{Conclusions}

In contrast to acceleration in direct current (DC) parallel electric fields
which in MHD simulations depends on the choice of the resistivity in the Ohms
law, we concentrate on the acceleration due to magnetic
gradient and curvature drift effects in the
cascading reconnection current sheet.
We found that both trapped in the magnetic islands and precipitating
electrons can be accelerated by the perpendicular magnetic
gradients and curvatures.
Trapped energetic electrons contribute to the formation of bright spots along
the current sheet trailing CMEs or eruptive filaments) as well as the flare
loop-top HXR radiation by their Bremsstrahlung.
Precipitating electrons, on the other hand, cause ribbons of UV and EUV
brightening in the solar chromosphere.

Whether an electron becomes trapped or precipitating depends on the
initial conditions (e.g., for precipitation, an electron should have a position
around X-points, velocity $> 2$ $v_{th}$ and pitch angle $\neq
90^{\circ}$).
Trapped electrons are energized mainly in the  magnetic islands in the coarse magnetic
fields,
while in the better resolved magnetic fields, the strongest trapped electron
acceleration takes place close to the X-points due to there finely resolved
larger magnetic curvatures and gradients in the smaller-scale magnetic
fields.
Both trapped and precipitating electrons are accelerated or decelerated in
dependence on their initial positions and pitch angles.
The electron final kinetic energy strength depends on the initial electron energy -
larger initial energies cause stronger accelerations.
As well as every kind of electron can get more energization if the smaller-scale
magnetic structures, obtained by higher resolution MHD simulations, are
taken into account.
Also because of these smaller-scale structures, energization of more
accelerated trapped electrons are mainly in the parallel direction in the finer
magnetic fields.
Other (less accelerated trapped and precipitating) electrons mainly gain
energies in the perpendicular direction.
Due to the asymmetry in the magnetic curvature drift acceleration term
around the center of 2.5D current sheet, the larger magnetic curvatures in the better
resolved magnetic structures cause stronger non-symmetric accelerations
around initial pitch angle $90^{\circ}$ of trapped and precipitating
electrons.
On the contrary, in the coarsely resolved magnetic fields both
trapped and precipitating electron acceleration
are close to symmetric around $90^{\circ}$.

With the better resolved small-scale magnetic structures, the maximum energy
gain of trapped electrons can be up to $421$ $keV$. This already suffices to
explain the observed loop-top HXR radiations.
Under the thin target model together with a simple Bethe-Heitler formula for
the cross section of Bremsstrahlung and an initial distributions
function $\propto E_{0}^{0}$ (or constant) and $\propto E_{0}^{-3}$ ,
the HXR spectral indices of trapped electrons can be as hard as $\sim5$
in the better resolved magnetic fields.
This is already hard enough to explain the observed HXR spectra in medium
solar flares.
For initial Maxwell-Boltzmann distributions for $T=10^6$ $K$, the HXR spectra
provide, however, just a slight enhancement of the high energy tail.

In the chromospheric ribbon-shape locations of precipitating electrons,
electrons starting lower in the solar atmosphere precipitate closer to the PIL.
The weak parallel acceleration of precipitating electron leads electrons
with initial pitch angles $<90^{\circ}$ precipitate to one side of the PIL,
while ones with initial pitch angles $>90^{\circ}$ go to the other side of the PIL.
Generally, there is a anti-symmetrical geometry of precipitating electron
locations in chromosphere around the PIL.
While because of the stronger accelerations of electrons with initial pitch
angles $>90^{\circ}$, more energetic electrons are located in
one side of the PIL only with the better resolved smaller-scale magnetic
structures.

\subsection{Discussion}

Solar flare observations imply that a large number of energetic electrons
should precipitate into the solar chromosphere where they cause observable
radiations.
Our calculations have shown that only $12 \%$ electrons can precipitate within
$10$ $t_{0}$.
While the whole current sheet evolution is as long as 520 $t_{0}$.
Depending on the magnetic field evolutions (see panels of
Fig.\ref{wholeView}), lower magnetic islands
($x < 70$ $L_{0}$) in Fig.\ref{EP_Trapped}) will merge into one
magnetic loop (see right panel of Fig.\ref{wholeView}) eventually.
So in the end, the electrons previously trapped in the lower magnetic islands
can later also precipitate to the chromosphere.
Taking into account this merging effects, more than
$63\%$ electrons will finally reach the chromosphere.
As well as when the space scale collapses to the kinetic one, the guiding center
approximation will be not valid any more. Particle motion will become
chaotic due to nonlinear resonances between particle bounce motion
and gyration. With the transition to chaos, \citealp{Buechner1989}
found that trapped nonadiabatic charged particles  can escape
due to chaotic pitch angle scattering effects.
Furthermore in this study during $10$ $t_{0}$, the background electromagnetic
fields are constant, so the time effects on electron acceleration are
neglected. With the evolutions of the electromagnetic fields, maybe
some trapped electrons become precipitating ones.
As a result, even more electrons will precipitate.

Our study can explain the observed medium and small solar flare loop-top HXR
spectra and EUV-ribbons just based on magnetic gradient and curvature effects in
magnetic islands without ad hoc postulated "anomalous" resistivity.
Precipitating electrons in our results, however, cannot explain the HXR
spectral indices in the foot-points
of solar flares which can be as hard as 1.5 in large solar flares.
Precipitating electrons also could reach the energies necessary to explain the
observed HXR spectra bases on magnetic gradient and curvature effects if
more and smaller magnetic field structures are formed by cascading magnetic
reconnection.

\appendix
          \section{Validation of the accuracy by using the conservation of the Second Adiabatic Invariant}
          \label{new_0_Fled}

\begin{figure}[htbp]
      \centering
          \mbox{
                \includegraphics[width=0.45\textwidth]{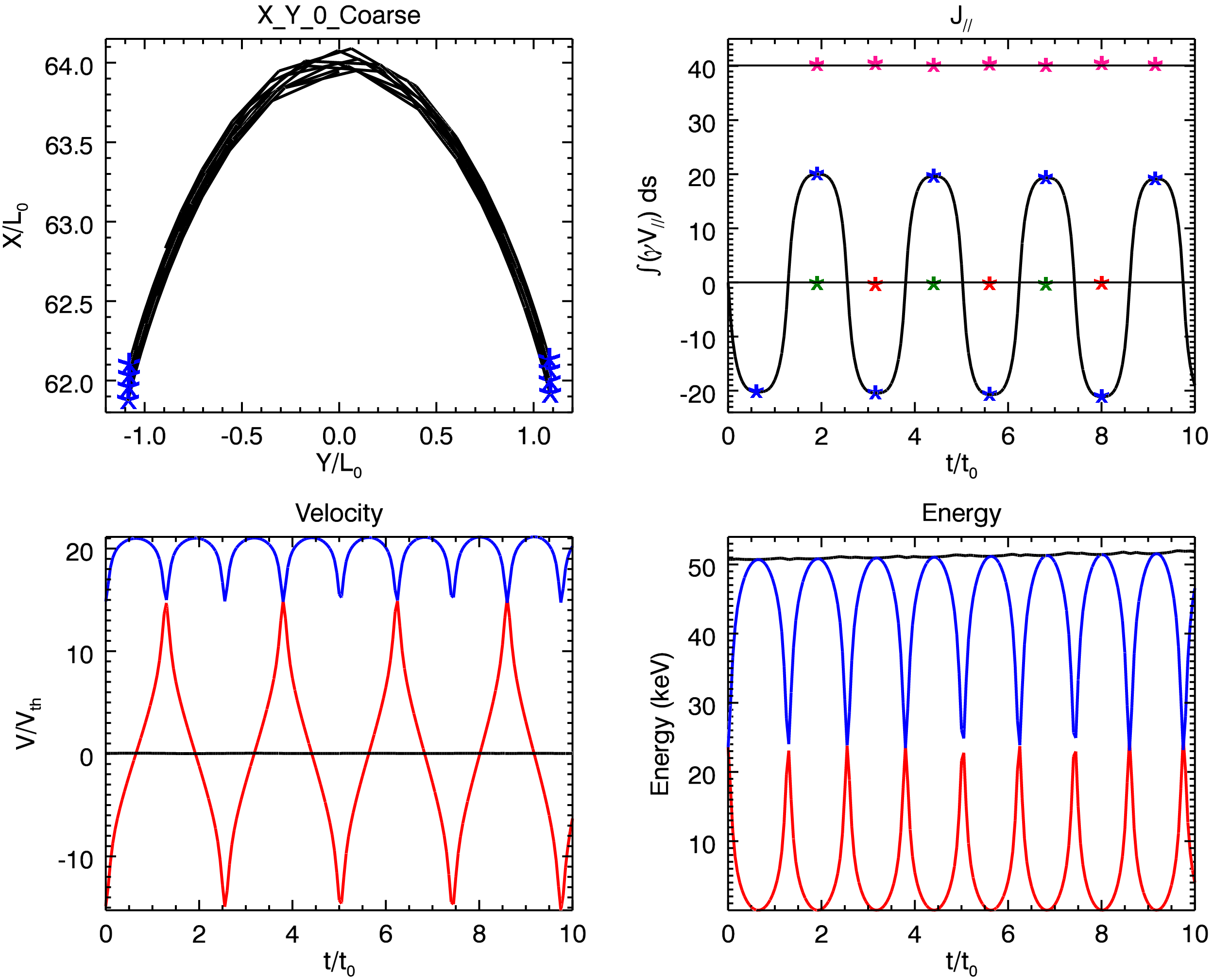}
               }
          \caption{Characteristic electron orbit indicating the conservation of the second adiabatic invariant.
                   The \textit{upper left} panel shows the XY-projection of the trajectory and the \textit{blue $\ast$} points in this panel
                   corresponding to the mirror points.
                   The \textit{upper right} panel shows $J_{\parallel}$ values along the trajectory and its \textit{red, green and deep pink $\ast$} points
                   show absolute changes of $J_{\parallel}$ during half-period (deep pink $\ast$ points) or one-period
                   (red and green $\ast$ points).
                   Three kinds of velocity (perpendicular gyration velocity - \textit{blue} line, parallel velocity - \textit{red} line and
                   drift velocity - \textit{dark} line )
                   and energy (perpendicular gyration energy - \textit{blue} line, parallel energy - \textit{red} line and
                   total kinetic energy - \textit{dark} line) are
                   separately shown in the \textit{bottom left} and \textit{right} panel.
                  }
          \label{Adiabatic}
      \end{figure}

The conservation of the second adiabatic invariant (\citealp{Northrop1963})
of trapped electrons can be used to validate the accuracy of the numerical
scheme solving Eqs.\eqref{GC1} to \eqref{GC4}:
\begin{align}
J_{\parallel} = m \int^{a}_{b} v_{\parallel} dl \cong \texttt{constant}
\label{A2}
\end{align}
In Eq.\eqref{A2}, the integral is taken along the particle guiding center
trajectory between the mirror points 'a' and 'b'.
Fig.\ref{Adiabatic} shows an example electron with conserved $J_{\parallel}$.

The upper left panel of Fig.\ref{Adiabatic} depicts the XY-projection of the
electron trajectory.
Every blue asterisks in the upper right panel of Fig.\ref{Adiabatic}
corresponds to a mirror point.
Note that in the definition of the second adiabatic invariant $J_{\parallel}$
(Eq.\eqref{A2}), $dl > 0$ when the particle is moving
forward and $dl < 0$ when it is moving backward.
For a convenience here we do not change the sign of 'dl' when
calculating the values of $J_{\parallel}$ along the electron trajectory.
One easily obtains not only the constant $J_{\parallel}$ over half-period
(deep pink '$\ast$' points) but also the vanishing $J_{\parallel}$ over a whole
bounce period (red and green '$\ast$' points) as one can see in the upper right
panel of Fig.\ref{Adiabatic}.
A comparison of parallel and perpendicular gyration velocities with the drift
velocities along the electron trajectory (bottom left panel of Fig.\ref{Adiabatic})
illustrates the validation of the second adiabatic invariant.
The bottom right panel of Fig.\ref{Adiabatic} indicates that the total
kinetic energy is exchanged between the parallel and the perpendicular
directed motion.

%
\begin{acknowledgements}
      This work was supported by the German-Chinese center of
      collaboration by the Max-Planck-Institute for Solar System Research,
      by the Max-Planck-Princeton Center for Plasma Physics and by the
      program of 2011CB811402 from MSTC and grants of 11233008, 11427803 from NNSFC.
      Lots of thanks to all members (e.g., Jan Skala, Patricio Munoz,
      Fabien Widmer) of the Theory and Simulation of Solar System Plasmas
      (TSSSP) group and Thomas Wiegelmann
      in the Max Planck Institute for Solar System Research (MPS), Germany for
      their useful and fruitful discussions and suggestions.

\end{acknowledgements}



\bibliographystyle{aa}
\bibliography{Reference}


\end{document}